\documentclass[12pt,aps,pre,amsmath,amssymb]{revtex4}
\usepackage{xcolor}
\usepackage{setspace}
\usepackage{graphicx}
\usepackage{amssymb}
\usepackage{amsmath}
\usepackage{mathrsfs}
\usepackage{calligra}
\usepackage{bm}
\usepackage{bbold}
\usepackage{hyperref}
 
\newcommand{\pfrac}[2]{ \frac{\partial #1}{\partial #2} }  

\newcommand{\w}{{\rm w}}

\newcommand{\e}{\mathrm e}
\renewcommand{\d}{\mathrm d}
\newcommand{\ii}{\mathrm i}

\newcommand{\rr}{ {\bf r} }

\newcommand{\kk}{ {\bf k} }
\newcommand{\RR}{ {\bf R} }

\newcommand{\D}{ {\cal D} }
\newcommand{\lB }{ l_{\rm B} }

\newcommand{\np}{n_{\rm p}} 

\newcommand{\gmm}[1]{{ g^{\rm mm}_{#1} }}
\newcommand{\gmc}[1]{ { g^{\rm mc}_{#1}}}
\newcommand{\gcc}[1]{{ g^{\rm cc}_{#1}}}

\begin{document}

$\null$
\hfill Revised November 2021\\
$\null$
\hfill {\footnotesize{(Typographical errors}}
{\footnotesize{corrected and references updated August 13, 2022;}}\\
$\null$
\hfill {\footnotesize{Fig.~8 corrected August 30, 2022)}}\\
$\null$ \hfill First Submitted May 2021
\vskip 0.3in

\begin{center}
{\Large\bf Numerical Techniques for Applications of Analytical}\\ 

\vskip 0.3cm

{\Large\bf Theories to Sequence-Dependent Phase Separations}\\ 

\vskip 0.3cm

{\Large\bf of Intrinsically Disordered Proteins}\\

\vskip .5in
{\bf Yi-Hsuan L{\footnotesize{\bf{IN}}}},$^{1,2,\dagger\P}$
{\bf Jonas W{\footnotesize{\bf{ESS\'EN}}}},$^{1,\dagger}$
{\bf Tanmoy P{\footnotesize{\bf{AL}}}},$^{1,\dagger}$
{\bf Suman D{\footnotesize{\bf{AS}}}},$^{1}$
 and\\
{\bf Hue Sun C{\footnotesize{\bf{HAN}}}}$^{1,*}$

$\null$

$^1$Department of Biochemistry,
University of Toronto, Toronto, Ontario M5S 1A8, Canada;\\
$^2$Molecular Medicine, Hospital for Sick Children, Toronto, 
Ontario M5G 0A4, Canada\\

\vskip 1.3cm

%

\end{center}

\vskip 0.3cm

\noindent
$\dagger$ Contributed equally to this work.
\\

\noindent
$\P$ Present address: HTuO Biosciences Inc., Vancouver, B.C., Canada
\\

\noindent
$*$Corresponding author\\
{\phantom{$^\dagger$}}
E-mail: {\tt huesun.chan@utoronto.ca};
Tel: (416)978-2697; Fax: (416)978-8548\\
{\phantom{$^\dagger$}}
URL:
\href{http://biochemistry.utoronto.ca/person/hue-sun-chan/}
{\tt http://biochemistry.utoronto.ca/person/hue-sun-chan/}\\
{\phantom{$^\dagger$}}
Mailing address:
{\phantom{$^\dagger$}}
Department of Biochemistry, University of Toronto,
Medical Sciences\\
{\phantom{$^\dagger$}}
Building -- 5th Fl.,
1 King's College Circle, Toronto, Ontario M5S 1A8, Canada.\\

\centerline{\bf {-----------------------------------------------------------------------------------------------------------}}

$\null$

\centerline{To appear as Chapter 3 in a 
{\href{https://www.barnesandnoble.com/w/phase-separated-biomolecular-condensates-huan-xiang-zhou/1141485430}{\it Methods in Molecular Biology}} 
volume ($\sim$ October 2022) entitled}
\centerline{\href{https://link.springer.com/book/9781071626627}
{``Phase-Separated Biomolecular Condensates''}}

$\null$
\centerline{{\it Edited by} Huan-Xiang Zhou, Jan-Hendrik Spille, 
and Priya Banerjee}
\centerline{Tentative page numbers of this chapter in the volume: pages~51--93}

\begin{center}
{\tt (hyperlinks to codes and related information are provided
in the\\.pdf version of this preprint)}
\end{center}

\vfill\eject

\noindent
{\large\bf Abstract}\\

\noindent
Biomolecular condensates, physically underpinned to a significant
extent by liquid-liquid phase separation (LLPS), are now widely 
recognized by numerous experimental studies to be of fundamental 
biological, biomedical, and biophysical importance. In the face of 
experimental discoveries,
analytical formulations emerged as a powerful yet tractable tool 
in recent theoretical investigations of the role of LLPS in the 
assembly and dissociation of these condensates. The pertinent LLPS often 
involves, though not exclusively, intrinsically disordered proteins 
engaging in multivalent interactions that are governed by their amino acid 
sequences. For researchers interested in applying these 
theoretical methods, here we provide a 
practical guide to a set of
computational techniques devised for extracting sequence-dependent 
LLPS properties from analytical formulations.  The numerical procedures
covered include those for the determinination of spinodal 
and binodal phase boundaries from a general free energy function with
examples based on the random phase approximation in polymer 
theory, construction of tie lines for multiple-component LLPS, and 
field-theoretic simulation of multiple-chain heteropolymeric systems using 
complex Langevin dynamics. Since a more accurate physical picture 
often requires comparing analytical theory against explicit-chain model 
predictions, a commonly utilized methodology for 
coarse-grained molecular dynamics simulations of sequence-specific LLPS 
is also briefly outlined.
\\

$\null$\\


\vfill\eject

\noindent{\Large\bf 1 Introduction}\\

Biomolecular condensates in living organisms are functional microsopic 
compartments that, despite lacking a lipid-membrane boundary, maintain
higher concentrations of specific sets of biomolecules within 
themselves compared with those of their surroundings~\cite{rosen2017}.
Because of this defining feature, micron-sized intracellular biomolecular 
condensates are often referred to as membraneless organelles.
A major physico-chemical driving force for the assembly of these condensates 
is the multivalent favorable interactions among biomolecules, resulting
in liquid-liquid phase separation (LLPS), with higher concentrations of 
specific biomolecular species in the condensed phase~\cite{CellBiol,NatPhys}.
In this regard, a biomolecular condensate may be likened to a liquid droplet.
The biomolecules involved can encompass intrinsically disordered proteins 
(IDPs), proteins containing both folded domains and intrinsically disordered 
regions (IDRs), globular proteins, and nucleic acids, or a subset thereof.
As a versatile form of spatio-temporal organization of biochemical 
processes through the microenvironments they create, biomolecular 
condensates serve critical physiological functions. Accordingly, their 
dysfunction or misregulation can lead to disease (reviewed, e.g., 
in refs.~\cite{cliff2017,Monika2018Rev}). While
functional intracellular biomolecular condensates can contain hundreds of 
biomolecular species organized in an extremely complex~\cite{tjian}, 
out-of-equilibrium~\cite{dresden} manner,
recent experimental advances indicate that a wealth of information on the 
biophysical properties and biological functions of biomolecular condensates 
can already be gleaned from equilibrium phase behaviors of highly simplified 
cell-free model systems consisting of one or only a few key biomolecular 
species as in the cases of FUS~\cite{McKnight12} and Ddx4~\cite{Nott15}. 
Therefore, it stands to reason that fundamental physical insights into how 
properties of biomolecular condensates are determined by the amino acid 
sequences of their major components, i.e., how molecular recognition leading
to the assembly of biomolecular condensates is effectuated, 
can be gained by computational modeling and 
theoretical considerations of simplified heteropolymeric LLPS 
systems~\cite{biochemrev,Roland2019}.
\\


\noindent{\bf 1.1 Theoretical constructs at different levels of realism}

With this perspective in mind, this chapter focuses primarily
on analytical theories for sequence-dependent IDP/IDR LLPS. Analytical 
theories and computational models of heteropolymer LLPS may be constructed 
at different levels of structural and energetic realism~\cite{biochemrev},
with a general trade-off between realism and computational 
efficiency. At one extreme, polymer LLPS can be described at a basic, 
mean-field level by Flory-Huggins (FH) theory \cite{Flory1953} for polymers 
with spatially short range, contact-like interactions and Overbeek-Voorn (OV)
theory~\cite{Overbeek1957} for polymers with spatially long range Coulomb 
interactions. The formalism of these theories are simple and the necessary
numerical calculations for determining their predicted LLPS properties 
are straightforward. However, for IDP/IDR LLPS, 
while FH and OV theories account approximately 
for effects of sequence composition (the 
numbers of different types of amino acid residues in the IDP/IDR sequence),
fundamentally FH and OV are not equipped to fully account for sequence 
dependence because they make no distinction between sequences that 
share the same amino acid composition.

At the other extreme, simulations of atomistic IDP/IDR models using an 
explicit representation of water molecules should offer a much more realistic 
picture of LLPS enriched with structural and energetic 
details \cite{regis}; but the computational 
efficiency of atomistic modeling of IDP/IDR LLPS is a major current
challenge because it requires simulating a large number of protein chains.
Recent advances in this direction include an explicit-water, atomistic 
molecular dynamics study of IDP condensed-phase properties that relies and
is built upon an initial implicit-solvent coarse-grained chain simulation 
for achieving equilibrium LLPS of the model system~\cite{ZhengetalJPCB2020}.
Because of the high computational cost of explicit-water atomic simulation, 
the number of LLPS systems that the research community can afford to examine 
by a fully atomistic approach would likely remain quite limited 
in the foreseeable future.
\\

\noindent{\bf 1.2 Random phase approximation in polymer theory}

In this context, analytical theories and computational models that
embody intermediate realism yet offer a higher degree of computational 
tractability are valuable for basic conceptual advances because they
allow for efficient exploration of extensive arrays of possible 
physical scenarios. One of such approaches is the random phase 
approximation (RPA) in polymer theory, which goes beyond mean-field theories 
such as the FH and OV formalisms, which neglect fluctuations of polymer 
concentration, by an approximate account that takes into consideration 
the lowest-order concentration fluctuations \cite{Ermoshkin2003}.
The most significant progress enabled by RPA theories is their ability
to account approximately for sequence dependence, as exemplified by 
the first theoretical account~\cite{Lin2016,Lin2017a} of the 
substantial difference in experimentally observed LLPS propensity 
between wildtype and a charge scrambled variant of Ddx4 IDR that share 
the same amino acid composition \cite{Nott15}.

RPA theories of sequence-dependent LLPS and related theoretical
developments were instrumental in uncovering \cite{Lin2017b}
and elucidating \cite{Alan} the relationship between isolated single-chain
IDP properties \cite{Lin2017b} or binding affinity of a pair of 
IDPs \cite{Alan} on one hand and multiple-chain LLPS 
propensity on the other \cite{Lin2017b,Alan,jeetainPNAS}.
An application of RPA to the LLPS of systems with two IDP species first
indicated that their populations in the condensed phase are miscible
when the sequence charge patterns \cite{rohit2013,kings2015} of the two 
IDP species are similar, but
they tend to demix into subcompartments when their sequence charge
patterns are significantly different \cite{Lin2017c}. The robustness of
this prediction has since been buttressed by explicit-chain 
coarse-grained molecular dynamics as well as field-theoretic
simulations \cite{Pal2021}. This finding highlights a stochastic, ``fuzzy''
type of molecular recognition that may contribute to subcompartmentalization
of some biomolecular condensates \cite{Lin2017c} such as the nucleolus.
The FIB1 and NPM1 IDPs that demix in the condensed phase of a simple cell-free
model experimental system for the nucleolus \cite{feric} do have very
different sequence charge patterns \cite{Lin2017c}, although in-cell
subcompartmentalization of the nucleolus likely involves much more 
complicated, non-equilibrium mechanisms \cite{cliff_2021_rev}.
More recently, the development of an improved version of RPA with Kuhn length 
renormalization---referred to as rG-RPA---has further extended the 
applicability of RPA theories of sequence-dependent LLPS beyond that of neutral 
or near-neutral polyampholytes to include strongly charged polyelectrolytes 
as well \cite{Lin2020}.

Comparisons between predictions from analytical RPA theories and results 
from simulations of more realistic coarse-grained chain models conducted
thus far indicate that the general trend of sequence dependence
stipulated by the two approaches to modeling LLPS are consistent,
but RPA tends to predict higher LLPS propensities than those predicted
by coarse-grained explicit-chain models \cite{DasLattice2018,DasPCCP2018}.
This limitation of RPA, which is likely caused in large measure by its 
highly idealized treatment of chain excluded volume, should be borne in 
mind when RPA predictions are interpreted vis-\`a-vis experimental data 
or results from molecular dynamics simulations.
\\

\noindent{\bf 1.3 Polymer field-theoretic simulation}

A more accurate---though computationally more intensive---method to extract 
observable consequences from typical polymer-theory Hamiltonians (energy 
functions) is field-theoretic simulation 
(FTS) \cite{Fredrickson2006,McCartyJPCL2019,joanPNAS}.
Unlike RPA which accounts only for the lowest order of concentration
fluctuations, in principle FTS is capable of accounting for 
concentration fluctuations in their entirety, though FTS is limited
nonetheless by the necessary discretization and finite size of the simulation
system. A comparison of FTS and RPA results on sequence-dependent
polyampholyte LLPS indicates reassuringly that they are largely 
consistent except RPA is seen to be inaccurate, as expected, for a
relatively small concentration regime where polymer is very 
dilute in solution \cite{McCartyJPCL2019}. 

FTS has been applied to compute correlation functions for systems 
with two IDP species. The results indicate that sequence charge-pattern 
mismatch and a strong generic excluded volume repulsion are both necessary 
for two polyampholytic IDP species to demix, i.e.,
to subcompartmentalize, in the condensed phase \cite{Pal2021}.
This conclusion is in line with the aforementioned RPA prediction with regard
to sequence charge pattern mismatch because
a strong excluded volume was tacitly enforced in that prior formulation by 
an incompressibility constraint \cite{Lin2017c}.
The finding underscores the sensitivity of FTS predictions to
excluded-volume assumptions, urging caution in choosing
and interpreting FTS parameters for excluded volume \cite{Pal2021}.
Insofar as electrostatic effects in polyampholyte LLPS are concerned,
recent studies using FTS, RPA, and coarse-grained explicit-chain 
molecular dynamics of polyampholytes with a simplified explicit model 
of polar solvent suggest that a reduced relative permittivity contributed 
by the solvent in the condensed phase probably leads only to a small to 
moderate enhancement of LLPS propensity compared to that predicted by 
an implicit-solvent model that assigns a uniform relative permittivity
corresponding to that of the bulk solvent throughout the entire
system volume \cite{DasPNAS2020,WessenJPCB2021}. In light of this
result, further quantitative
details of this basic comparison between explicit- and implicit-solvent
formulations should be pursued using atomistic models.
\\


\noindent{\bf 1.4 Software for theoretical studies of sequence-dependent LLPS}

This chapter is not a comprehensive review of theoretical and computational
approaches to biomolecular condensates. This chapter is intended 
as a practical guide for readers who are interested
in applying the above-described theoretical methods, serving as an introduction
to the computer codes we have developed to perform the pertinent numerical 
calculations.
Accordingly, the presentation of analytical formulations below is minimal; and
we will refer readers to the published literature for much of the
mathematical details whenever they are available. The codes written by us for 
LLPS investigation and cited in the discussion below
are available from the following GitHub software sharing webpages:

\noindent
\href{https://github.com/laphysique/Protein_RPA}
{\tt https://github.com/laphysique/Protein{\textunderscore}RPA}

\noindent
\href{https://github.com/laphysique/FH_LLPS_simple_system}
{\tt https://github.com/laphysique/FH{\textunderscore}LLPS{\textunderscore}simple{\textunderscore}system}

\noindent
\href{https://github.com/jwessen/IDP_phase_separation}
{\tt https://github.com/jwessen/IDP{\textunderscore}phase{\textunderscore}separation}

\noindent
\href{https://github.com/mmTanmoy/IDP_phase_separation}
{\tt https://github.com/mmTanmoy/IDP{\textunderscore}phase{\textunderscore}separation}

\noindent
\href{https://github.com/laphysique/FTS_polyampholyte_water}
{\tt https://github.com/laphysique/FTS{\textunderscore}polyampholyte{\textunderscore}water}

\noindent
which may also be accessed via the links on our group's webpage for LLPS
software:

\noindent
\href{https://arrhenius.med.utoronto.ca/~chan/llps_software.html}
{\tt https://arrhenius.med.utoronto.ca/{\textasciitilde}chan/llps{\textunderscore}software.html}
\\

\noindent{\bf 1.5 Coarse-grained explicit-chain simulation and other approaches}

In addition to analytical theories, a very brief summary is also provided
below for a coarse-grained, explicit-chain molecular dynamics 
methodology for studying amino acid sequence-dependent LLPS \cite{Dignon2018}
that relies on a protocol for efficient equilibration by initializing 
simulation with a highly condensed polymer slab in an elongated 
rectangular simulation box \cite{Silmore2017}. Among the method's many 
applications (e.g., refs.~\cite{jeetainPNAS,Schuster2020}), its results
have been utilized---as mentioned above---to assess 
RPA \cite{DasLattice2018,DasPCCP2018} and FTS \cite{Pal2021}
predictions as well as to evaluate \cite{DasPNAS2020}
common coarse-grained LLPS potentials \cite{Dignon2018} for their
adequacy in describing $\pi$-related interactions in LLPS \cite{robert}
and IDP interactions in general \cite{kaw2013,cosb15}.

Besides the methods mentioned above, it should be noted that 
other theoretical/computational approaches, including
lattice \cite{lassi,anders2020,stefan2019}, 
patchy particle \cite{hxzhou2018,Cam_patchy}, 
and restricted primitive \cite{Sing2017,SingPerry2020} models, have 
provided important insights to biomolecular condensates as well; but they
are beyond the scope of this chapter.
\\

\noindent{\bf 1.6 Step-by-step practical guides to software}

In the presentation below, numerous references will be made to the specific 
codes available via the webpages listed in Sect.~1.4 above for readers who
are interested in test-running them to facilitate understanding of the
theoretical formulations and/or to use the codes with appropriate
adjustments to perform computations for their own research efforts.
For beginners in the field, we have compiled several additional
``Practical guides'' sections in the discussion below to provide more 
detailed, step-by-step recipes for obtaining the data necessary 
to produce some of the figures in this chapter. These should serve
as useful, concrete examples of how the codes work in practice.
\\


\noindent
{\Large\bf 2 Construction of Phase Diagrams from Analytical\\ 
{\phantom{2\ }}Free Energy Functions}\\


Analytical theories for LLPS allow for numerical efficiency when 
the free energy of a given system, $f$, is provided as a function of the 
concentrations $\rho$'s of the constituents:
\begin{equation}
f = f(\rho_1, \rho_2, ... \rho_n, \rho_\w) \; .
\label{eq:f0}
\end{equation}
Here, as is customary in the polymer theory literature, the
free energy $f$ is given in units of $Vk_{\rm B}T$, where $V$ is system 
volume, $k_{\rm B}$ is Boltzmann's constant, and $T$ is absolute temperature.
As such, $f$ is a Helmholtz free energy defined for a system with
constant $V$ rather than a Gibbs free energy defined for a system
under constant pressure $P$. Nonetheless, Helmholtz and Gibbs free 
energies for biomolecular LLPS under ambient conditions
are expected to be similiar because of relatively small $PV$ contributions to 
Gibbs free energy under those conditions. It should be noted, however, 
that $PV$ contributions can be substantial for biomolecular LLPS under 
high hydrostatic pressures as experienced in the deep sea~\cite{Roland2019}.
We do not consider high-$P$ LLPS in this chapter.

In Eq.~\ref{eq:f0}, $\rho_1,\rho_2,\dots,\rho_n$ are the concentrations for
$n$ solute species, and $\rho_{\rm w}$ is solvent concentration.
A subscript ``w'' is used for solvent because water is the most common solvent 
in biological systems. The $\rho$'s may be given as number densities or
volume fractions. When the $\rho$'s are volume fractions (note that
volume fractions are often denoted by $\phi$'s in the literature), 
$(\sum_{i=1}^n \rho_i) +\rho_\w = 1$ by definition.

In such theories, the system with a single phase is macroscopically 
homogeneous. Local fluctuations in $\rho$'s are effectively taken into 
account by analytical methods, mostly via various perturbative approaches,
such that an $f$ in the form of Eq.~\ref{eq:f0} may be arrived at
without intensive numerical calculation.
These analytical theories include, but are not limited to:
\begin{itemize}
        \setlength\itemsep{0em}
\item Flory-Huggins theory~\cite{Flory1953},
\item Overbeek-Voorn theory~\cite{Overbeek1957},
\item Flory-Stockmayer gelation theory~\cite{Flory1953,Semenov1998},
\item Polymer cluster theory~\cite{Bawendi1987, Baker1993},
\item Wertheim's thermodynamic perturbation theory (TPT)~\cite{Wertheim1986,Kastelic2016},
\item Random-phase approximation (RPA) theory~\cite{Ermoshkin2003, Lin2016, Lin2017a, Lin2017b, Lin2017c},
and
\item Renormalized-Gaussian fluctuation / random-phase-approximation (RGF/rG-RPA) theory~\cite{Lin2020,ZGWang}.
\end{itemize}

We now provide a guide to the basic general procedure for extracting 
information about phase behaviors from 
$f(\rho_1, \rho_2, ... \rho_n, \rho_\w)$
in the form of phase diagrams, irrespective of how 
$f(\rho_1, \rho_2, ... \rho_n, \rho_\w)$ is derived.
\\

\noindent{\bf 2.1 Spinodal decomposition}

The impact of fluctuations in solvent and solute concentrations on
the stability of the system may be examined by the following Taylor series 
expansion of the free energy $f$ around any set of concentrations
$\{\rho_i\}\equiv\{\rho_1,\rho_2,\dots,\rho_n, \rho_\w\}$ (note
that the index $i$ now covers the solvent as well, i.e., $i=1,2,\dots,n,\w$):
\begin{equation}
f(\{\rho_i+\delta\rho_i\}) = f(\{\rho_i\}) 
+ \sum_{i}\frac{\partial f}{\partial \rho_i}\delta\rho_i 
+ \frac{1}{2}\sum_{i,j}\frac{\partial^2 f}{\partial \rho_i \partial \rho_j}\delta\rho_i \delta\rho_j + O(\delta\rho^3) \; ,
\label{eq:Taylor}
\end{equation}
where the derivatives are evaluated at $\{\rho_i\}$.
The system is deemed thermodynamically stable if 
the partition function of the fluctuations, i.e., the integral
of $\exp(-f)$ over the $\delta\rho_i$'s, is finite. Considering terms
for $f$ in Eq.~\ref{eq:Taylor} through second order in $\delta\rho_i$'s while
recognizing that the contribution from the $f$ term linear in $\delta\rho_i$ 
to $\exp(-f)$ may be absorbed into a term quadratic in $\delta\rho_i$ 
in the argument for the exponential by the usual procedure of linear
shifting of integration variables for Gaussian integrals, 
this condition for stability amounts to requiring
the Hessian matrix $\hat{\cal H}$, with matrix elements
\begin{equation}
\hat{\cal H}_{ij} =  \frac{\partial^2 f}{\partial \rho_i \partial \rho_j} \; , 
\label{eq:HM}
\end{equation}
to be positive definite---meaning that
$\hat{\cal H}$ is symmetric (which follows automatically
from Eq.~\ref{eq:HM} for any smooth, real-number function $f$) 
and that all its eigenvalues are positive ($>0$). 
Physically, the all-positive-eigenvalues condition ensures that
the homogeneous solution phase is stable against any 
$\{\delta\rho_i\}$ fluctuations in solute and solvent concentrations.
Spinodal decomposition occurs when at least one of the
eigenvalues of $\hat{\cal H}$ becomes zero, at which point 
the determinant of the matrix vanishes, i.e.,
\begin{equation}
\det \hat{\cal H} = 0 \; .
\label{eq:deteq}
\end{equation}
Conversely, $\hat{\cal H}$ being not positive definite implies global 
instability of the homogeneous solution phase,
manifested experimentally by spinodal decomposition.
It should be emphasized, however, that although $\det \hat{\cal H} = 0$ defines
the spinodal boundary and that $\det \hat{\cal H}<0$ always indicates that
there is at least one negative eigenvalue and therefore the homogeneous
solution phase is not stable, $\det \hat{\cal H}>0$, in contrast,
does not guarantee that the homogeneous
solution phase is thermodynamically stable~\cite{Lin2017c}.
The homogeneous solution phase is not stable
when $\hat{\cal H}$ has one or more negative eigenvalues, 
because it is unstable against fluctuations along any of the eigenvectors 
associated with those negative eigenvalues. 
However, while an odd number of negative eigenvalues leads to 
$\det \hat{\cal H}<0$, an even number of negative eigenvalues results 
in $\det \hat{\cal H}>0$. Hence, instability of the homogeneous solution 
phase can, in some cases, be concomitant with $\det \hat{\cal H}>0$.
In other words, $\det \hat{\cal H}<0$ is sufficient but not necessary
for spinodal decomposition.

In the case of incompressible two-component systems that satisfy
$\rho + \rho_{\rm w} = 1$ (without loss of generality, the unit for 
concentration is chosen such that the maximum concentration is unity, 
see example in Fig.~1), there is only one independent concentration
variable and thus the Hessian matrix $\hat{\cal H}$ reduces to a single 
second derivative. It follows that the boundary condition for spinodal 
instability, $\det \hat{\cal H}=0$, takes the form of
\begin{equation}
f''(\rho) \equiv  \frac{\partial^2 f}{\partial \rho^2} = 0 \; .
\label{eq:spinodal0}
\end{equation}
When $\rho$ is allowed to vary under some---but not all---fixed
environmental conditions (temperature, hydrostatic pressure, 
etc. \cite{Roland2019}), there are two $\rho$'s 
satisfying the spinodal boundary condition Eq.~\ref{eq:spinodal0}, one 
on the dilute (relatively smaller-$\rho$) 
side and the other on the condensed (relatively larger-$\rho$) side of the free 
energy function (Fig.~1a, open circles).  When environmental conditions 
such as temperature that affect the interaction 
strength (symbolized as $u$) are varied, the free energy function $f$ also 
varies accordingly. At a certain interaction strength $u_{\rm cr}$ at
which the $\rho$ values for the two spinodal boundaries 
merge into a single $\rho=\rho_{\rm cr}$,
the $u_{\rm cr}$ and $\rho_{\rm cr}$ are recognized as the critical interaction
strength and critical concentration (hence the subscript ``cr''), respectively.
These two critical quantities are determined by the equations
\begin{subequations}
\begin{align}
f''(\rho_{\rm cr}, u_{\rm cr}) 
\equiv & \left.\frac{\partial^2 f}{\partial \rho^2}\right|_{\rho=\rho_{\rm cr}, u=u_{\rm cr}} = 0\\
f'''(\rho_{\rm cr}, u_{\rm cr}) 
\equiv & \left.\frac{\partial^3 f}{\partial \rho^3}\right|_{\rho=\rho_{\rm cr}, u=u_{\rm cr}} = 0 \; .
\end{align}
\end{subequations}
The interaction strength $u$ is characterized by different parameters
in different models. For instance, in FH theory, $u$ corresponds to
the $\chi$ parameter; in RPA for electrostatics, $u$ can be quantified
by the Bjerrum length $l_{\rm B}$.

The critical interaction strength $u_{\rm cr}$ may be obtained through 
minimizing the function $u=u(\rho)$ defined implicitly by
the spinodal condition $f''(\rho, u(\rho)) = 0$ because $u_{\rm cr}$ 
is the {\it weakest} interaction strength that allows for
phase separation.
Numerically, $u(\rho)$ can be determined by a root-finding algorithm such 
as Newton's method or bisection method. Subsequently, 
$(\rho_{\rm cr}, u_{\rm cr})$ can be solved by minimization algorithms 
such as the Nelder-Mead method~\cite{Nelder1965} or BFGS 
method~\cite{Roger1987}.
A pseudocode for this procedure is as follows:

\begin{center}
\begin{minipage}{0.70\textwidth}
{\tt

def crictial\_point():

$\qquad$rho\_c, u\_c = minimization(u\_spinodal, var=rho)

$\qquad$return rho\_c, u\_c

\ \

def u\_spinodal(rho):

$\qquad$u\_sol = root\_find(ddf, var=u, const=rho)

$\qquad$return u\_sol

}
\end{minipage}

\ \

\end{center}

Once the critical point is located, it is guaranteed that 
for all $u > u_{\rm cr}$, there is only one solution for 
$f''(\rho)=0$ in the $(0, \rho_{\rm c})$ range and one solution
in the $(\rho_{\rm c}, 1)$ range; 
and bisection root-finding algorithms can be applied to
locate the two spinodal concentrations.

Regarding the interaction strength $u$, it should be noted that while
$u=u(T)\propto 1/T$ is often assumed in idealized models with a
temperature-independent energy scale ${\cal E}$ such that $u(T)\sim {\cal E}/T$
(the Bjerrum length $l_{\rm B}$ is an example of this type of $u$),
the $T$-dependence of interaction strengths of the aqueous solvent-mediated 
interactions underlying biomolecular condensates is often more complex.
In those cases, instead of being a constant, ${\cal E}$ depends on $T$, i.e.,
${\cal E}\rightarrow {\cal E}(T)$, and
$u(T)$ is not necessarily a monotonic function of $T$ (unlike $u\propto 1/T$).
For instance, the $u(T)$ for hydrophobic interaction has a maximum
at $T_{\rm max}\approx 25^\circ$C (see, e.g., the $\chi(T)$ function in
Fig.~2 of ref.~\cite{Dill1989}).
This means that a hydrophobicity-driven LLPS may entail two $u_{\rm cr}$ 
values, one corresponding to a higher critical temperature,
$T_{\rm cr}^{\rm UCST}$, which is an upper critical solution temperature 
(UCST), and the other a lower critical temperature, $T_{\rm cr}^{\rm LCST}$, 
which is a lower critical solution temperature 
($T_{\rm cr}^{\rm UCST}>T_{\rm cr}^{\rm LCST}$) \cite{biochemrev,Roland2019,jeetainACS}.
The physical reason behind this phenomenon is that
LLPS is impossible for $T>T_{\rm cr}^{\rm UCST}$ because configuration
entropy overwhelms favorable interchain interaction at high temperatures; 
and LLPS is also impossible for $T<T_{\rm cr}^{\rm LCST}$ because of 
the weakening of hydrophobic interactions at low temperatures.
When performing numerical calculations for model systems embodying this 
feature, one has to treat the region in which $u(T)$ increases with
increasing $T$ and the region in which $u(T)$ decreases with increasing $T$
separately so that the mapping from $u$ to $T$, i.e., the inverse
function $T = u^{-1}(u(T))$, can be defined.  
\\

\begin{figure}[ht!]
   \centering
   \includegraphics[width=0.65\textwidth]{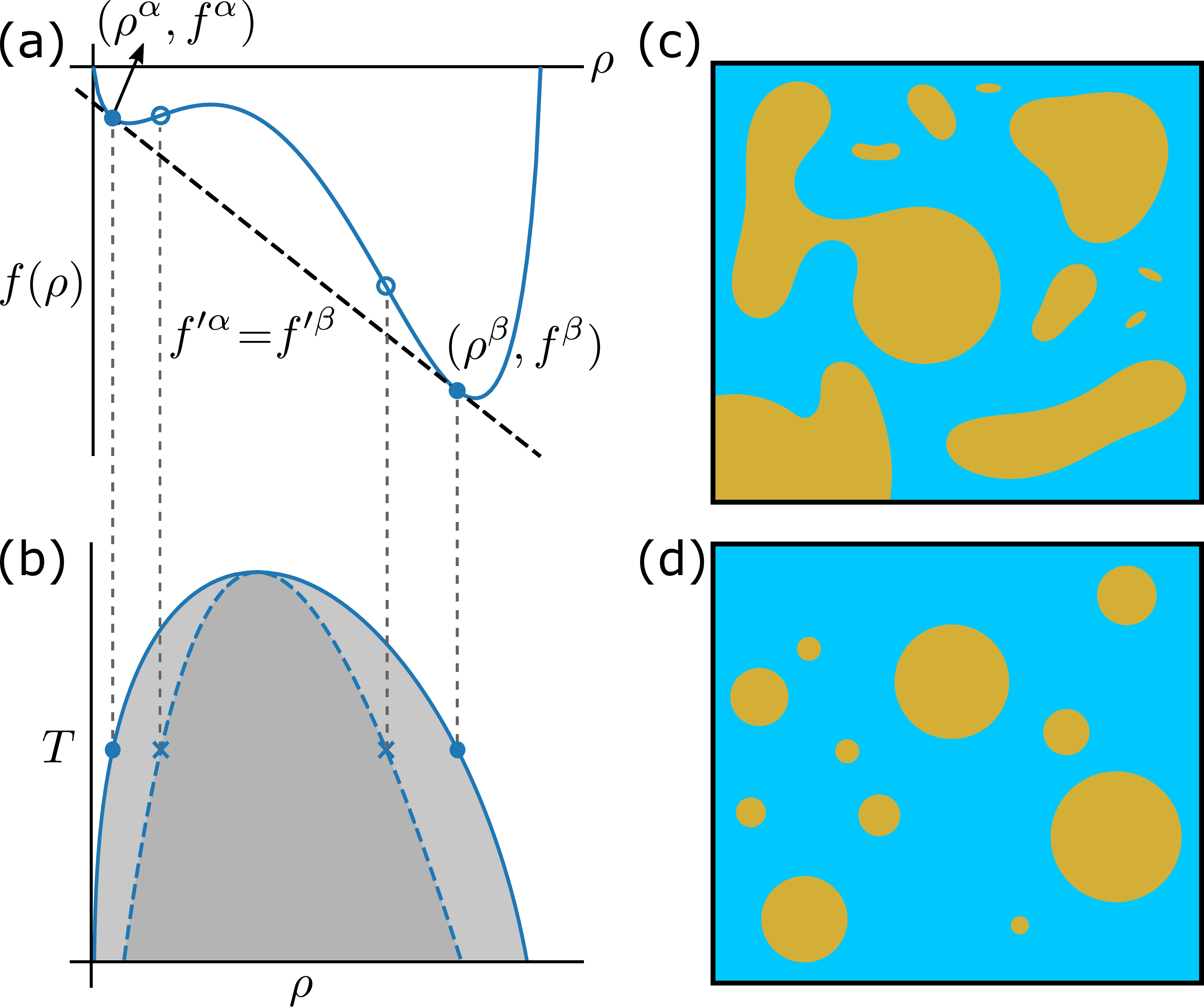}
\caption{Construction of the spinodal and binodal (coexistence) phase
boundaries. An illustrative example is shown for the procedure described in 
the text and in ref.~\cite{Lin2017a}.
(a) Free energy in units of $Vk_{\rm B}T$, $f(\rho)$, is given
by the blue curve for one selected temperature.
The $f''(\rho)=0$ inflection points for the spinodal phase boundary
(Eq.~\ref{eq:spinodal0}) are marked by the two open circles. 
The common tangent (inclined dashed line) of the two coexistence 
phases for binodal phase separation is determined by 
the two equalities in Eq.~\ref{eq:com_tan} to obtain the coexisting
concentrations $\rho^\alpha$ and $\rho^\beta$ marked by the two filled circles. 
(b) The spinodal (dashed blue curve) and binodal (solid blue curve) phase
boundaries are constructed by joining the $\rho$ values of
the open and filled circles in (a), respectively, at different
temperatures, as indicated by the vertical dashed lines between (a) and (b).
Here $T$ is absolute temperature. 
The shaded regions represent the spinodal (darker) 
and binodal (lighter) regimes of the phase diagram.
Results in (a) and (b) are those of a FH model with
$f(\rho)=(\rho/N)\ln\rho+(1-\rho)\ln(1-\rho) + \chi\rho(1-\rho)$,  
$N=3$, and $T=1/\chi$ (same as that in Fig.~1 of ref.~\cite{biochemrev});
the $f(\rho)$ curve in (a) is for $\chi=1.7$.
(c) and (d): Cartoons of spinodal decomposition (c) 
and droplet formation in binodal phase separation (d), wherein
solute is depicted in gold and solvent in cyan.
Microscopic images of IDP spinodal decomposition and binodal phase separation 
can be found, e.g., for Ddx4 in Fig.~7 of ref.~\cite{Lin2017a} and 
Fig.~3 of ref.~\cite{Nott15} respectively.
}
   \label{fig:common_tangent}
\end{figure}


\noindent{\bf 2.2 Coexisting phases}

The spinodal condition signifies divergences 
in concentration fluctuations and global breakdown of solution homogeneity.
In contrast, when the interaction strength of a system is gradually increased
in experiments, the phase separation state that is first observed
is composed of multiple {\it coexisting} thermodynamically {\it stable}
phases in which local concentration fluctuations do not diverge.
This phenomenon is referred to as binodal phase separation.

The coexisting phases satisfy a balance requirement under which the chemical
potentials for each type of solute across all phases are identical, and 
the osmotic pressure (which depends on the chemical potential 
of solvent)~\cite{Lin2017a} is also identical across all phases. 
In other words, in a solution of $n$ solute species that
separate into $m$ coexisting phases, 
\begin{equation}
\mu_i^{\alpha} = \mu_i \; , \quad \Pi^\alpha= \Pi,
\end{equation}
where $\mu$ denotes chemical potential and $\Pi$ denotes 
osmotic pressure; $i$, $\alpha$ are labels, respectively, for solute 
species and phases; $\mu_i$ is the chemical potential 
common to solute type $i$ across all phases (same for all $\alpha$).
Similarly, $\Pi$ is the common osmotic pressure across all phases 
(again, same for all $\alpha$).

By definition, the chemical potential of solute $i$ in phase $\alpha$ 
is given by
\begin{equation}
\mu_i^\alpha = f'^{\alpha}_i \equiv 
\left.\pfrac{f}{\rho_i}\right|_{\rho_i = \rho_i^\alpha}
\; ,
\label{eq:coexist1}
\end{equation}
and the osmotic pressure in phase $\alpha$ is given by
\begin{equation}
\Pi^\alpha = f^{\alpha} - \sum_{i=1}^n \rho_i^\alpha f'^{\alpha}_i
\; ,
\label{eq:coexist2}
\end{equation}
where the free energy 
$f^{\alpha} \equiv 
f(\rho_1^\alpha, \rho_2^\alpha , \cdots, \rho_n^\alpha, \rho_\w^\alpha)$.
For binary phase separation in an incompressible two-component ($n=1$) system
with $\rho_1 + \rho_{\rm w} = 1$,
there is only one independent concentration variable $\rho_1$. 
Accordingly, the balance requirement for coexisting phases $\alpha$, $\beta$ 
can be simplified by combining the general conditions in
Eqs.~\ref{eq:coexist1} and \ref{eq:coexist2} and using $\rho$ to denote 
solute concentration $\rho_1$ (as in Eq.~\ref{eq:spinodal0}) to yield 
\begin{equation}
f'^{\alpha} = f'^\beta = \frac{f^\alpha - f^\beta}{\rho^\alpha - \rho^\beta}
\; ,
        \label{eq:com_tan}
\end{equation}
which is the {\it common tanget} condition illustrated in Fig.~1a.

Mapping the coexisting (binodal) condition and the
spinodal condition (Sect.~2.1) in Fig.~1a onto
a solute concentration-temperature plot yields the phase diagram in Fig.~1b.
Cartoons of spatial variation of concentration in spinodal and binodal
phase separations are provided in Fig.~1c and d.

Besides the above consideration of chemical potential and osmotic pressure,
the balance requirment for coexisting phases can also be derived
by minimizing the system free energy,
\begin{equation}
f_{\rm sys} = \sum_{\alpha=1}^m v^\alpha f^\alpha \; ,
\label{eq:fsys1}
\end{equation}
under the constraints of volume and mass conservation, viz.,
\begin{equation}
\sum_{\alpha=1}^m v^\alpha = 1 \; , \quad \sum_{\alpha=1}^m v^\alpha \rho_i^\alpha = \rho_i^{\rm bulk} 
\; ,
\label{eq:fsys2}
\end{equation}
where $v^\alpha$ is the fractional volume in phase $\alpha$, and 
$\rho_i^{\rm bulk}$ is the overall concentration of solute $i$
averaged over the entire system volume.
As an example, for an incompressible two-component system,
$f^\alpha=f(\rho^\alpha)$, $f^\beta=f(\rho^\beta)$, thus, by
denoting $v^\alpha$ as $v$ and therefore $v^\beta=1-v$,
\begin{equation}
f_{\rm sys} =
f_{\rm sys}^{\rm 2comp} = v f(\rho^\alpha) + (1-v) f\left(\rho^\beta=\frac{\rho^{\rm bulk}-v\rho^\alpha}{1-v}   \right) \; .
        \label{eq:f_coexist_3comp}
\end{equation}
Because $f_{\rm sys}^{\rm 2comp}$ is a function of variables
$v$ and $\rho^\alpha$, the minimum of $f_{\rm sys}^{\rm 2comp}$ 
is determined by
\begin{equation}
\frac {\partial f_{\rm sys}^{\rm 2comp}}{\partial\rho^\alpha} 
= \frac {\partial f_{\rm sys}^{\rm 2comp}}{\partial v} =0 \; ,
        \label{eq:J_coexist_3comp}
\end{equation}
yielding
\begin{subequations}
\begin{align}
v f'^\alpha - v f'^\beta & = 0 \; , \\
f^\alpha - f^\beta + f'^\beta \left(\rho^\beta - \rho^\alpha \right) &=0 \; ,
\end{align}
        \label{eq:J_coexist_3comp_sep}
\end{subequations}
which are precisely the equations for solving the common 
tangent in Eq.~(\ref{eq:com_tan}).
\\


\noindent{\bf 2.3 Systems with three or more components}

For incompressible three-component systems satisfying 
$\rho_1 + \rho_2 + \rho_\w =1$, there are two independent concentration
variables. When $\rho_1$ and $\rho_2$ are chosen as independent variables,
it follows from Eq.~\ref{eq:deteq} that the boundary condition for 
the spinodal region is given by
\begin{equation}
\det \hat{\cal H} =
\left|
\begin{matrix}
\frac{\partial^2 f}{\partial \rho_1^2} & \frac{\partial^2 f}{\partial \rho_1 
\partial \rho_2} \\[1em]
\frac{\partial^2 f}{\partial \rho_2 \partial\rho_1} & \frac{\partial^2 f}{\partial \rho_2^2}
\end{matrix}
\right| = 0 \; .
\label{eq:2component-spinodal}
\end{equation}
To exemplify how this condition is applied, Fig.~2 provides the free 
energy surface (Fig.~2a) of a simple three-component system, 
its spinodal region (Fig.~2b and c),
and the initial steps for determining the region 
(Fig.~2b). As shown by the 25 grid 
points in Fig.~2b, the spinodal region can be determined approximately 
by a grid search for the area with $\det \hat{\cal H}<0$. 
In practice, a denser grid should be used than that shown in Fig.~2b to 
arrive at a better-approximated spinodal region, which can then be 
used as a stepping stone to arrive at a much more precise spinodal region 
by utilizing the numerical techniques described below.  
Formulations of three-component LLPS have recently been applied
to study systems with two polyampholyes with different charge
patterns (as model IDPs)~\cite{Lin2017c} and models of natural IDP 
solutions with salt~\cite{Lin2020}. In this connection, it is worth noting
that short-chain (small $N$) FH models similar to those used in Fig.~1 
and Fig.~2 for illustration can be useful for modeling LLPS of folded
proteins as well~\cite{mz1,mz2}.
A step-by-step practical guide to the calculations in these simple
FH models are discussed in Sect.~2.5 below.

\begin{figure}[ht!]
   \centering
   \includegraphics[width=0.98\textwidth]{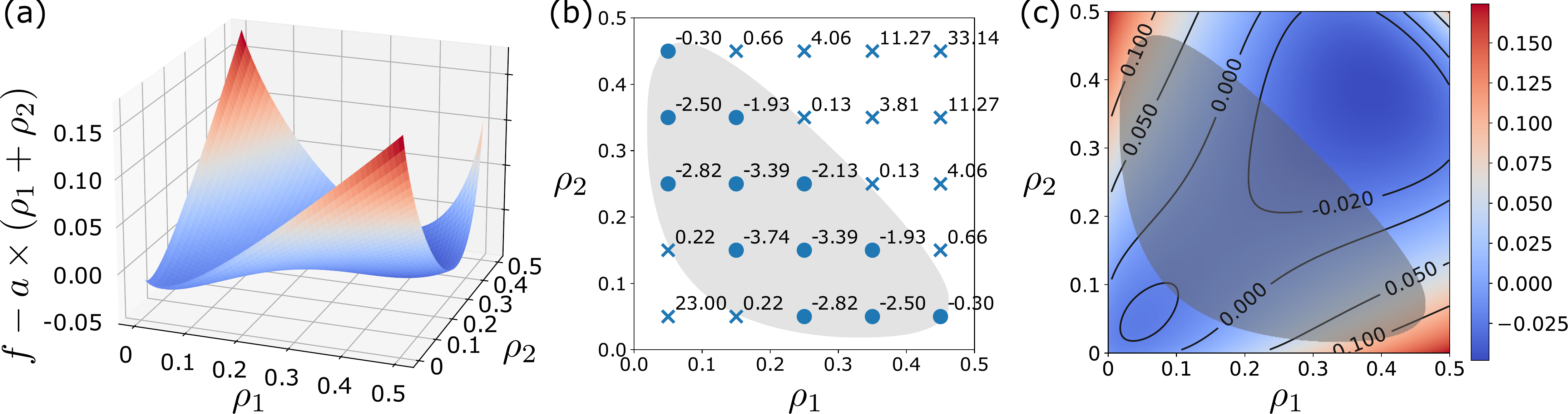} 
   \caption{Three-component phase separation.
Shown here is an example free energy surface 
of a solution containing two solute species with 
overall concentrations $\rho_1$ and $\rho_2$ in a solvent,
as a three-dimensional plot (a) with 
its spinodal region (b) and as an equivalent contour plot (c).
Free energy $f$ in (a) and (c) is color 
coded by the scale in (c).
The region shaded in grey in (b) and (c) is the spinodal area defined by
the boundary condition in Eq.~\ref{eq:2component-spinodal}.
This example is a FH model with 
$f(\rho_1,\rho_2) = (\rho_1/N)\ln\rho_1 + (\rho_2/N)\ln\rho_2 + 
(1-\rho_1-\rho_2)\ln (1-\rho_1-\rho_2) - 
\chi[\rho_1\rho_2  + 0.3\rho_1(1-\rho_1) + 0.3\rho_2(1-\rho_2)]$ and
$N=3$, where $\chi=1.7$ is used.
As in Fig.~2a of ref.~\cite{Lin2017c}, a term linear in $\rho_1$ and $\rho_2$ 
is included in the variable for $f$ to highlight variation in 
curvature of the $f(\rho_1,\rho_2)$ function. Such a shift in free energy 
has no effect on phase separation. The same free energy variable with 
$a=-0.72505$ is used throughout this figure.
To illustrate how the spinodal region is determined computationally,
the $\det \hat{\cal H}$ (Eq.~\ref{eq:2component-spinodal}) 
values at 25 example grid points 
on the $\rho_1$---$\rho_2$ plane are provided in (b). The grid
points are depicted as circles when $\det \hat{\cal H}<0$
and as crosses when $\det \hat{\cal H}>0$.
We have also conducted an extensive numerical check to verify that the 
Hessian matrix $\hat{\cal H}$ has two positive eigenvalues in the 
$\det \hat{\cal H}>0$ region of this example.
}
   \label{fig:two_component}
\end{figure}

In studies of three-component systems, phase behaviors are often
provided by two-dimensional phase diagrams with solute concentrations 
$\rho_1$ and $\rho_2$ as variables at a given interaction strength.
To construct such phase diagrams, one may start with the spinodal region,
which can be determined by sampling a grid on the $\rho_1$--$\rho_2$ plane
to identify the $(\rho_1, \rho_2)$ values on the grid for which
$\det \hat{\cal H} < 0$, 
as illustrated above in Fig.~2b.
Another example is now provided in Fig.~3a.
With a sufficiently dense grid,
this is an efficient way to identify the approximate spinodal region. As
will be explained below, this approximate spinodal region is important 
as a starting point to efficiently map out the coexisting region as well.
To arrive at an accurate spinodal boundary beyond the rough boundary
afforded by the grid search, we seek to determine the spinodal boundary
as a mapping between $\rho_1$ and $\rho_2$ by numerically solving
the pertinent mathematical relations.
To facilitate this determination, we first separate the spinodal
region into an upper part (with larger $\rho_2$ values) and a lower
part (with smaller $\rho_2$ values) with a demarcation
at or near the $\rho_2$ value for which the spinodal region's extent 
in the $\rho_1$ direction is the widest (Fig.~3b, horizontal dashed line) 
so that most likely $\rho_1$ is mapped onto only one value of $\rho_2$ in 
each of the two parts of the spinodal region. This step is necessary since
it allows us to construct the spinodal boundary as two separate mathematical 
functions $\rho_2(\rho_1)$, one for the upper part and one for the lower part 
of the spinodal region. Otherwise, in the absence of this separation of
the spinodal region, one $\rho_1$ value is always mapped onto 
two $\rho_2$ values (and vice versa) when
the entire spinodal boundary is considered as a whole, in which case 
a mathematical function in the form of $\rho_2(\rho_1)$ is not defined 
($\rho_1$ maps onto mutiple values of $\rho_2$). 
Now, $\rho_2(\rho_1)$ can be solved separately for the two parts by
applying the pseudocode {\tt u{\textunderscore}spinodal(rho)}
mentioned in Sect.~2.1 above by treating $\rho_2$ as $u$ and
substituting $\rho_1$ for the variable {\tt rho}.
As an illustration, this procedure is used to obtain the
grey-shaded spinodal area for the example free energy surface in Fig.~2b.

For coexisting phases (binodal phase separation), the same correspondence 
between balancing of chemical potentials and free energy
minimization for binary phase separation of an incompressible
three-component system is described in ref.~\cite{Lin2017c} (see,
in particular, Eqs.~11--22 of this reference). 
For systems with more components and more separated phases, 
the form of $f_{\rm sys}$ as defined by Eqs.~\ref{eq:fsys1} 
and \ref{eq:fsys2} can be too complicated for algebraic minimization 
algorithms. In those cases, more numerically intense algorithms such 
as Monte Carlo methods may be applied~\cite{Jacobs2017}.

In numerical calculations, the common tangent equation in Eq.~(\ref{eq:com_tan})
can be solved via multidimensional root-finding algorithms, such as Broyden's
method~\cite{Broyden1965} or modified Powell's method~\cite{Press2007}; the
minimization of $f_{\rm sys}$ can be solved by multidimensional minimization
algorithms such as L-BFGS-B~\cite{Byrd1995} and SLSQP~\cite{Kraft88} methods.
It should be noted that multidimensional root-finding and minimization 
algorithms are often sensitive to the initial guess of the solutions. In 
many cases, a random search for a proper initial guess is required to 
arrive at the correct roots or global minimum.
Therefore, it would be helpful if the numerical search can be 
constrained to a smaller regime. With this in mind, it is noteworthy 
that when interaction strength or solute concentration of a homogeneous
solution is varied, binodal phase separation---with a mechanism akin to
nucleation---ensues {\it before} spinodal instability, suggesting that
the spinodal boundary is always
enclosed by the coexisting boundary (the spinodal region is a subset of the
region enclosed by the coexisting boundary).
Since determining the spinodal region is numerically efficient, the above
consideration implies that one can first determine the spinodal region and
then use the concentrations within the region---which is then known to
be unstable against phase separation---as input to numerical
algorithms for solving the coexisting (binodal) phase boundary.
We find that this is indeed a cost-effective protocol.
An example of such a calculation is described in Fig.~3c and d.
\\

\begin{figure}[ht!]
   \centering
   \includegraphics[width=0.66\textwidth]{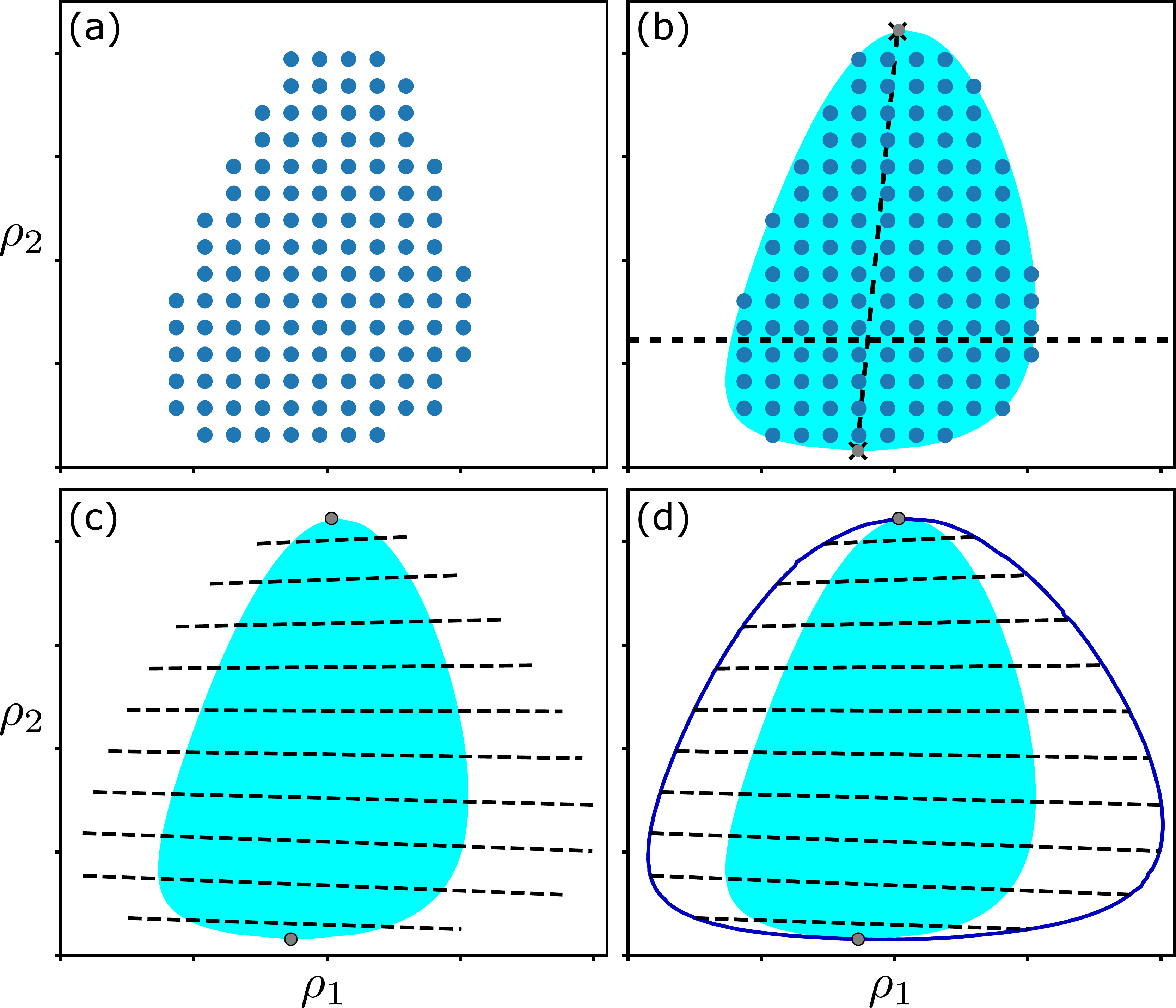}
   \caption{Construction of phase diagram of a three-component system 
with tie lines linking pairs of coexisting solute concentrations.
(a) Conduct a grid search to determine an approximate $\det \hat{\cal H} < 0$
spinodal region [marked by cyan shading in (b)--(d)].
(b) Choose a $\rho_2$ value separating the spindoal region into an upper and a 
lower parts at or near where the $\rho_1$-extent of the region is widest
(horizontal dashed line); then locate the maximum $\rho_2$ in the upper 
part and the minimum $\rho_2$ in the lower part (crosses)
on the $\det{\cal H} = 0$ boundary of the approximate spinodal region.
(c) Use points along the line linking the maximum and minimum of $\rho_2$ 
[near-vertical inclined dashed line in (b)] as initial bulk 
concentrations $(\rho_1^{\rm bulk}, \rho_2^{\rm bulk})$; 
calculate the final binodal phase-separated concentrations
$\alpha=(\rho_1^\alpha, \rho_2^\alpha)$ and $\beta = (\rho_1^\beta,
\rho_2^\beta)$ by minimizing system free energy using the prescription
for three-component phase separation described by Eqs.~11--22 of 
ref.~\cite{Lin2017c} (note that the three conditions in Eq.~22 of 
ref.~\cite{Lin2017c} is the three-component equivalent of the two
conditions in the two-component Eq.~\ref{eq:J_coexist_3comp_sep} in
this chapter).
Then connect $\alpha$, $\beta$ by dashed lines as tie lines.
The points marked by crosses in (b) are now marked by
circles in (c) and (d) as critical points.
   (d) Connect all $\alpha$ and $\beta$ points together with the upper 
and lower critical points to obtain the binodal phase boundary.
Data for this example are those of the theoretical rG-RPA phase 
diagram of Ddx4 at pH = 7.0 in Fig.~5a of ref.~\cite{Lin2020}.
An extensive numerical check has been conducted to verify that the 
$\hat{\cal H}$ is positive definite in the $\det \hat{\cal H}>0$ region 
of this example.
   }
   \label{fig:2d_contour}
\end{figure}


\noindent{\bf 2.4 Available software}

Some of the software we developed to perform numerical calculations 
for the RPA and rG-RPA theories mentioned above
are available from the webpage
\href{https://github.com/laphysique/Protein_RPA}
{\tt https://github.com/laphysique/Protein{\textunderscore}RPA}
of the Github repository.

For systems with one polymer species and salt at a given concentration 
(i.e., salt concentration is an input parameter, which can be zero or any
other chosen value, and treated as a constant during the execution of 
the code), the energy minimization solvers for common tangent construction 
are in
{\footnotesize
\href{https://github.com/laphysique/Protein_RPA/blob/master/f_min_solve_1p_1salt.py}
{\tt f{\textunderscore}min{\textunderscore}solve{\textunderscore}1p{\textunderscore}1salt.py}},
{\footnotesize
\href{https://github.com/laphysique/Protein_RPA/blob/master/f_min_solve_1p_1salt_LCST.py}
{\tt f{\textunderscore}min{\textunderscore}solve{\textunderscore}1p{\textunderscore}1salt{\textunderscore}LCST.py}}, 
and 
{\footnotesize
\href{https://github.com/laphysique/Protein_RPA/blob/master/fmin_rgRPA_1p_1salt.py}
{\tt fmin{\textunderscore}rgRPA{\textunderscore}1p{\textunderscore}1salt.py}} 
for different systems.
These scripts are used for plotting phase diagrams 
with temperature/interaction strength as variable.
A ``one button for all'' script for each of these systems is also available as
{\footnotesize
\href{https://github.com/laphysique/Protein_RPA/blob/master/ps_main_1p_1salt.py}
{\tt ps{\textunderscore}main{\textunderscore}1p{\textunderscore}1salt.py}}, 
{\footnotesize
\href{https://github.com/laphysique/Protein_RPA/blob/master/ps_main_1p_1salt_LCST.py}
{\tt ps{\textunderscore}main{\textunderscore}1p{\textunderscore}1salt{\textunderscore}LCST.py}}, and 
{\footnotesize
\href{https://github.com/laphysique/Protein_RPA/blob/master/rgRPA_main.py}
{\tt rgRPA{\textunderscore}main.py}}
to directly output the spinodal and binodal phase separation
concentrations in a temperature range for a given sequence and a salt
concentration. However, users will need to tune the parameters of these
``one button for all'' scripts to ensure that they work well for their polymer
system of interest.

For systems with one polymer species and variable salt concentration, 
the script 
{\footnotesize
\href{https://github.com/laphysique/Protein_RPA/blob/master/fsalt_rgRPA_1p_1salt.py}
{\tt fsalt{\textunderscore}rgRPA{\textunderscore}1p{\textunderscore}1salt.py}} 
is for solving polymer-salt two-dimensional phase diagrams in rG-RPA theory
(tracking changes in polymer as well as salt concentrations). 
This script has been applied in ref.~\cite{Lin2020}.
The corresponding ``one button for all'' script is
{\footnotesize
\href{https://github.com/laphysique/Protein_RPA/blob/master/rgRPA_main_salt.py}
{\tt rgRPA{\textunderscore}main{\textunderscore}salt.py}}.
In situations when the performance of this code is not robust, the
script
{\footnotesize
\href{https://github.com/laphysique/Protein_RPA/blob/master/rgRPA_salt_main_auto.py}
{\tt rgRPA{\textunderscore}salt{\textunderscore}main{\textunderscore}auto.py}} 
may be used to perform part of the task.  
For instance,
{\footnotesize
\href{https://github.com/laphysique/Protein_RPA/blob/master/rgRPA_salt_main_auto.py}
{\tt rgRPA{\textunderscore}salt{\textunderscore}main{\textunderscore}auto.py}},
which is relatively more robust, is employed to perform the calculation 
described in Fig.~3c and d, while scripts specialized to the system
of interest are needed to be written anew to conduct the grid search
and partition of the putative spinodal region as described in Fig.~3a and b.

Functions for the free energy minimization solver for two-polymer 
RPA system are in
{\footnotesize
\href{https://github.com/laphysique/Protein_RPA/blob/master/f_min_solve_2p_0salt.py}
{\tt f{\textunderscore}min{\textunderscore}solve{\textunderscore}2p{\textunderscore}0salt.py}}.
This code, which is used in ref.~\cite{Lin2017c}, is for obtaining 
two-dimensional phase diagrams with the concentrations of the two polymers 
as variables.  A ``one button for all'' script for outputing the entire
two-dimensional phase boundary is in
{\footnotesize
\href{https://github.com/laphysique/Protein_RPA/blob/master/ps_main_2p_0salt.py}
{\tt ps{\textunderscore}main{\textunderscore}2p{\textunderscore}0salt.py}}.
It should be noted, however, that the performance of these codes for 
two-polymer systems are sensitive to the precise form of the free energy 
function; the parameters in the codes will need to be tuned case by case.
\\

\noindent{\bf 2.5 Practical guide to the calculations for Figs.~1 and 2}

The specific codes for the FH models used in Figs.~1 and 2 are available
in the Jupyter Notebook format via the following Github webpage: 
{\footnotesize
\href{https://github.com/laphysique/FH_LLPS_simple_system/blob/main/FH_phase_diagram.ipynb}
{\texttt{https://github.com/laphysique/FH\_LLPS\_simple\_system/blob/main/FH\_phase\_diagram.ipynb
}}}
Note that this webpage provides only a static image because Github does 
not support dynamic Jupyter Notebook. Users will need to download the
file to their computer with the Jupyter Notebook package installed to 
run the code, or they may run Jupyter Notebook on Google Drive as follows:
(i) Upload the Notebook file
{\footnotesize
\href{https://github.com/laphysique/FH_LLPS_simple_system/blob/main/FH_phase_diagram.ipynb}
{\tt FH{\textunderscore}phase{\textunderscore}diagram.ipynb}} 
(obtained from the above Github website) to Google Drive.
(ii)  Double click it. A message ``No preview available'' will appear. 
Click the ``Open with'' button at the center top of the screen, then 
click ``Connect more apps''.
(iii) A window of ``Google Workplace Marketplace'' will show up. 
Scroll down and click ``Colaboratory'' and install it (Google will ask 
you to approve a few items).
(iv) After Colaboratory is installed, refresh the Google Drive page and double
click the Notebook. The Notebook will then run in a new browser window.

$\null$\\

\noindent
{\Large\bf 3 Field-Theoretic Simulation (FTS)}\\

FTS methods \cite{Fredrickson2006}, which are capable of modeling
sequence-dependent LLPS without invoking some of the approximations in
analytical theories such as RPA, are currently foraging in the fast 
expanding research territories of LLPS
in biology \cite{Pal2021,McCartyJPCL2019,joanPNAS,WessenJPCB2021,joanElife}.
As far as computational cost is concerned, FTS---in this respect it is similar 
to self-consistent field theory \cite{Matsen2006}---may be viewed as
intermediate between analytical theories such as RPA and explicit-chain
simulations. Compared to RPA and rG-RPA, significantly more intense
computation is required in FTS to 
calculate the chemical potentials $\mu$'s and osmotic pressure $\Pi$'s defined
in Eqs.~\ref{eq:coexist1} and \ref{eq:coexist2}.
Nonetheless, because the analytical formulation
of FTS reduces the tremendous number of coordinate degrees of freedom of 
the individual polymer chains and solvent molecules to relatively few 
density degrees of freedom, FTS incurs a significantly less 
computational cost vis-\`a-vis explicit-chain simulations---especially
those with explicit solvent molecules---which keep track of all coordinate 
degrees of freedom.
\\
 
\noindent{\bf 3.1 Field theory descriptions of polymer solutions}

Given the immense complexities of the biological systems, the
current focus of FTS model building is to capture the basic physics that drive
LLPS in living cells starting from \textit{minimal} interaction Hamiltonians
while making least amount of compromise on the accuracy in solving the
models. Minimality ensures least number of free parameters, thus offering firm
conceptual grasp on the physical factors involved.

To illustrate the usage of FTS, we consider a simple system of volume $V$
containing $\np$ polyampholytes, each consisting of $N$ beads connected via
Gaussian springs (extensively studied using FTS in 
ref.~\cite{McCartyJPCL2019}). On
every chain, bead $\alpha$ is assigned an electric charge $\sigma_{\alpha}$ in
units of the proton charge, and all beads are modelled as Gaussian
distributions $\Gamma(\rr)=\exp(-\rr^2/2 \bar{a}^2)/(2 \pi \bar{a}^2)^{3/2}$ with
a common width $\bar{a}$. This Gaussian smearing procedure was introduced in
ref.~\cite{Wang2010} to regulate ultraviolet (UV), i.e., short-distance, 
divergences associated with point interactions. 
For simplicity, we restrict the discussion in this Section to overall
electrically neutral chains, i.e.~$\sum_{\alpha=1}^N \sigma_{\alpha} = 0$.

The chains interact via Coulomb forces and short-range excluded volume
repulsion. To write down the partition function for this system, we first
express the bead and charge densities in the system as, respectively,
\begin{subequations}
\begin{align}
\hat{\rho}(\rr) &= \sum_{i=1}^{\np} \sum_{\alpha=1}^N \Gamma(\rr - \RR_{i,\alpha}) \, ,  \label{eq:micro_polymer_density} \\
\hat{c}(\rr) &= \sum_{i=1}^{\np} \sum_{\alpha=1}^N  \Gamma(\rr - \RR_{i,\alpha}) \sigma_{\alpha} \, ,
\end{align}
\end{subequations}
where $\RR_{i,\alpha}$ is the position vector 
of bead $\alpha$ on chain $i$. The canonical partition function is then
\begin{equation} \label{eq:Z_part}
Z = \frac{1}{\np!} \left( \prod_{i=1}^{\np} \prod_{\alpha=1}^N \int \d \RR_{i,\alpha} \right) \e^{- \hat{H}_0 - \hat{H}_1 - \hat{H}_2} \, ,
\end{equation}
where the three terms of the Hamiltonian, in units of $k_{\rm B}T$, 
\begin{subequations} \label{eq:H_part}
\begin{align}
\hat{H}_0 &= \frac{3}{2b^2} \sum_{i=1}^{\np} \sum_{\alpha=1}^{N-1} ( \RR_{i,\alpha+1} - \RR_{i,\alpha})^2  \, , \\
\hat{H}_1 &= \frac{v}{2} \int \d \rr \, \hat{\rho}(\rr)^2 \, , \\
\hat{H}_2 & = \frac{\lB}{2} \int \d \rr \int \d \rr' \, \frac{\hat{c}(\rr) \hat{c}(\rr') }{ | \rr - \rr' | } \, ,
\end{align}
\end{subequations}
provide, respectively, chain connectivity with $b$ being the reference 
bond length, excluded volume interactions with $v$ as the excluded volume 
parameter, and electrostatic interactions with Bjerrum length
$\lB = e^2 / 4 \pi \epsilon k_{\rm B} T$. 

A partition function such as Eq.~\ref{eq:Z_part} can be turned into the
partition function, $Z$, of a statistical field theory through the use of
Hubbard-Stratonovich transformations \cite{Edwards1965}. We refer to existing
literature for the detailed derivation \cite{Fredrickson2006,Fred2002} and 
simply state the result here:
\begin{equation} \label{eq:Z_field}
Z = \frac{ V^{\np} }{ \np! Z_w Z_{\psi} } \int \D w \int \D \psi \, \e^{-H[w,\psi] } \, ,
\end{equation}
where the field Hamiltonian is
\begin{equation} \label{eq:H_field}
H[w,\psi] = - \np \ln Q[\breve{w},\breve{\psi}] + \int \d \rr \left( \frac{w^2}{2 v} + \frac{(\bm{\nabla} \psi)^2}{8 \pi \lB} \right)
\; ,
\end{equation}
with
\begin{equation}
Q[\breve{w},\breve{\psi}] \equiv \frac{1}{V} \left( \frac{3}{2 \pi b^2} \right)^{\frac{3(N-1)}{2}} \left( \prod_{\alpha=1}^N \int \d \RR_{\alpha} \right) \exp\left[ - \frac{3}{2b^2}\sum_{\alpha=1}^{N-1}(\RR_{\alpha+1}-\RR_{\alpha} )^2 - \ii \sum_{\alpha=1}^N \left( \breve{w}(\RR_{\alpha}) + \sigma_{\alpha} \breve{\psi}(\RR_{\alpha}) \right) \right] 
\, ,
\end{equation}
wherein $\ii^2=-1$,
being the partition function of a single polymer subject to external chemical
potential and electrostatic potential fields 
$\ii \breve{w}(\rr) = \Gamma \star \ii
w(\rr) \equiv \int \d \rr' \Gamma(\rr - \rr') \ii w(\rr')$ and $\ii
\breve{\psi}(\rr) = \Gamma \star \ii \psi(\rr) \equiv \int \d \rr' \Gamma(\rr -
\rr') \ii \psi(\rr')$, respectively.  The renormalization factors $Z_w$ and
$Z_{\psi}$, given by
\begin{subequations}
\begin{align}
Z_w &= \int \D w \, \e^{ - \int \d \rr \, w^2 / 2 v }  \, , \\
Z_{\psi} &= \int \D \psi \, \e^{ - \int \d \rr \, (\bm{\nabla} \psi)^2 / 8 \pi \lB  } \, ,
\end{align}
\end{subequations}
are physically inconsequential, but are included for numerical convenience
since they cancel UV divergences associated with arbitrarily small fluctuations
in $w$ and $\psi$. This can be understood as follows: The Gaussian smearing
removes Fourier modes from $w$ and $\psi$ with wave numbers $k\equiv | \kk |  \gtrsim
\bar{a}^{-1}$, such that these modes are not present in
$Q[\breve{w},\breve{\psi}] $ in the field Hamiltonian in
Eq.~\ref{eq:H_field}. The UV parts of $\int \D w \int \D \psi \,
\e^{-H[w,\psi] }$ and $Z_{w} Z_{\psi}$ therefore exactly match, and therefore
cancel in the field theory partition function in Eq.~\ref{eq:Z_field}.
\\

\noindent{\bf 3.2 Complex-Langevin sampling}

The complex nature of $H[w,\psi]$ is problematic for many standard Monte Carlo
methods since $\exp(-H[w,\psi])$ cannot be interpreted as a probability weight
for a field configuration. This problem, known as the ``sign problem'', can be
tackled using Complex-Langevin (CL) sampling \cite{Parisi1983,Klauder1983}, 
which is closely related to the formulation for stochastic quantization 
in theories of quantum fields \cite{ParisiWu1981,ChanHapern86,Damgaard87}.
CL sampling
has been shown to constitute an efficient approach to polymer field theories
\cite{Fred2002}. The CL method introduces a fictitious time-dependence on the
fields, after analytically continuing them to their complex planes 
[$w(\rr) \rightarrow w(\rr,t)$ and $\psi(\rr) \rightarrow \psi(\rr,t)$ ],
that is governed by the stochastic differential equations 
\begin{equation}
\label{eq:CL_evolution}
\frac{\partial \phi(\rr, t)}{\partial t} = - \frac{\delta H[w,\psi]}{\delta \phi(\rr,t)} + \eta_{\phi}(\rr,t) \, , \quad \phi=w,\psi \, 
\end{equation}
for Langevin dynamics evolving in the fictitious time $t$,
where $\eta_{\phi}(\rr,t)$ represents real-valued Gaussian noise with zero mean
and $\langle \eta_{\phi}(\rr,t) \eta_{\phi'}(\rr',t') \rangle = 2
\delta_{\phi,\phi'} \delta( \rr - \rr' ) \delta( t - t') $. Averages in the
field picture are then computed as asymptotic CL time averages,
\begin{equation}
\label{eq:averages}
\langle {\cal O}[w(\rr),\psi(\rr)] \rangle_{\rm F} = 
\lim_{t_{\rm max} \rightarrow \infty} \frac{1}{t_{\rm max}} 
\int_0^{t_{\rm max}} \d t \, {\cal O}[w(\rr,t),\psi(\rr,t)] \; .
\end{equation}
For our model in Eq.~\ref{eq:H_part}, 
the functional derivatives of the field Hamiltonian are
\begin{subequations} \label{eq:H_derivatives}
\begin{align}
\frac{\delta H[w,\psi]}{\delta w(\rr)}  &= \ii \tilde{\rho}(\rr) + \frac{1}{v} w(\rr)   \, , \\
\frac{\delta H[w,\psi]}{\delta \psi(\rr)}       &= \ii \tilde{c}(\rr)- \frac{1}{4 \pi \lB} \bm{\nabla}^2 \psi(\rr)  \, ,
\end{align}
\end{subequations}
where
\begin{subequations}
\begin{align}
\tilde{\rho}(\rr) &= \ii \np \frac{\delta \ln Q[\breve{w},\breve{\psi}] }{\delta w(\rr) }   \, , \\
\tilde{c}(\rr)          &= \ii \np \frac{\delta \ln Q[\breve{w},\breve{\psi}] }{\delta \psi(\rr) }   \, 
\end{align}
\end{subequations}
are field operators corresponding to bead and charge densities, named so
because $\langle \hat{\rho}(\rr) \rangle = \langle \tilde{\rho}(\rr)
\rangle_{\rm F}$ and $\langle \hat{c}(\rr) \rangle = \langle \tilde{c}(\rr)
\rangle_{\rm F}$, where, as defined above in
Eq.~\ref{eq:averages}, $\langle\dots\rangle_{\rm F}$ stands for
field average \cite{Fredrickson2006}. For a given field configuration
$\lbrace w(\rr), \psi(\rr) \rbrace$, so-called forwards and backwards chain
propagators $q_F(\rr,\alpha)$ and $q_B(\rr, \alpha)$ can be used to
calculate the field operators $\tilde{\rho}(\rr)$, $\tilde{c}(\rr)$ and
$Q[\breve{w},\breve{\psi}]$. The chain propagators are constructed iteratively
as \begin{subequations}
\label{eq:propagators}
\begin{align}
q_F(\rr,\alpha+1) &= \e^{-\ii W(\rr,\alpha+1) } \Phi \star q_F(\rr,\alpha)   \, , \\
q_B(\rr,\alpha-1) &= \e^{-\ii W(\rr,\alpha-1)  } \Phi \star q_B(\rr,\alpha) \, ,
\end{align}
\end{subequations}
where $W(\rr,\alpha)\equiv \breve{w}(\rr) + \sigma_{\alpha}
\breve{\psi}(\alpha)$, $\Phi(\rr) \equiv 
(3 / 2 \pi b^2)^{3/2} \exp(-3 \rr^2 / 2 b^2)$, and
starting from $q_F(\rr,1) = \exp[-\ii W(\rr,1) ]$ and 
$q_B(\rr,N) = \exp[-\ii W(\rr,N)]$. One can then show that 
\begin{subequations}
\label{eq:Q_rhos}
\begin{align}
Q[\breve{w},\breve{\psi}] &= \frac{1}{V} \int \d \rr \, q_F(\rr,N)  \, , \\
\tilde{\rho}(\rr) &= \Gamma 
\star \frac{n_{\rm p}}{V \, Q[\breve{w},\breve{\psi}] } 
\sum_{\alpha=1}^{N}  \e^{\ii W(\rr,\alpha) } q_F(\rr,\alpha) 
q_B(\rr,\alpha) \, , \\
\tilde{c}(\rr) &= \Gamma \star \frac{n_{\rm p}}{V\, Q[\breve{w},\breve{\psi}] } 
\sum_{\alpha=1}^{N}  \e^{\ii W(\rr,\alpha) } q_F(\rr,\alpha) 
q_B(\rr,\alpha) \sigma_{\alpha} \, .
\end{align}
\end{subequations}

In FTS, the continuous fields are approximated by discrete field variables
defined on the sites of a cubic lattice with periodic boundary conditions. As
noted above, field fluctuations on scales $\lesssim \bar{a}$ can be
``renormalized away'' which means that the lattice spacing $\Delta x$ should be
chosen such that $\Delta x \lesssim \bar{a}$. The CL time evolution equations
in Eq.~\ref{eq:CL_evolution} are integrated numerically using a finite
time-step $\Delta t $, where $\Delta t \, \eta_{w}(\rr,t)$ and $\Delta t \,
\eta_{\psi}(\rr,t)$ are random numbers drawn independently for each lattice
site and time-step from a normal distribution with zero mean and variance
$\sigma^2 =  2 \Delta t /  {\Delta x}^3 $.

A simple Euler stepping method to solve Eq.~\ref{eq:CL_evolution} is
possible, but often requires an extremely small time-step $\Delta t$ to yield
numerical stability. The numerical issue is reminiscent of a similar issue
in dealing with stiff
differential equations, which often can be handled by implicit integration
schemes. Several sophisticated numerical methods have been proposed for
integrating the CL evolution equations in Eq.~\ref{eq:CL_evolution} that
yield improved numerical stability \cite{Lennon_etal_2008}. We here review the
first order semi-implicit method of ref.~\cite{Lennon_etal_2008}, generalized to
multiple fields $\lbrace \phi_i \rbrace$. Since a fully implicit integration,
where $\delta H / \delta \phi_i$ would be evaluated at $t+\Delta t$, is not
possible due to the complicated dependence of $H$ on the fields, the
semi-implicit method instead works by replacing the linear part of $\delta H /
\delta \phi_i$ by its value at $t + \Delta t$, i.e.,  
\begin{equation}
\label{eq:semi_implicit_def}
\phi_i(t+\Delta t) = \phi_i(t) + \Delta t \left[ - \left( \frac{\delta H}{\delta \phi_i} \right)_{t} + \left( \left[ \frac{\delta H}{\delta \phi_i} \right]_\mathrm{lin} \right)_{t} - \left( \left[ \frac{\delta H}{\delta \phi_i} \right]_\mathrm{lin} \right)_{t+\Delta_t} \right] +\Delta t \, \eta_i  
\; ,
\end{equation}
where $[\dots]_{\rm lin}$ denotes taking the linear part in fields.
We have here suppressed the spatial dependence for notational convenience.
Provided that the quadratic expansion of the field Hamiltonian is
expressed in the form
\begin{equation}
\label{eq:Field_hamiltonian_expansion}
H = \int \d \rr \left( \sum_i a_i \phi_i + \frac{1}{2} \sum_{i,j} \phi_i K_{ij} \phi_j + \mathcal{O}(\phi^3) \right) \, 
\end{equation}
with $K_{ij}$ to be determined below,
the linear part of $\delta H / \delta \phi_i$ is
\begin{equation}
\left( \left[ \frac{\delta H}{\delta \phi_i} \right]_\mathrm{lin} \right)_{t} = a_i + \sum_j K_{ij} \phi_j(t) \, .
\end{equation}
Solving Eq.~\ref{eq:semi_implicit_def} for $\phi_i(t+\Delta t)$ gives
\begin{equation}
\label{eq:semi_implicit_update}
\phi_i(t+\Delta t) = \phi_i(t) +  \mathcal{F}^{-1} \sum_j  \left[ (\mathsf{M}^{-1})_{ij} \, \mathcal{F} \left[ -\Delta t \left( \frac{\delta H}{\delta \phi_j} \right)_{t} + \Delta t \, \eta_j\right] \right] \, ,
\end{equation}
where $\mathcal{F}$ and $\mathcal{F}^{-1}$ denote Fourier and inverse Fourier transforms, respectively, and $\mathsf{M}^{-1}$ is the matrix inverse of the matrix $\mathsf{M}$ whose components are given by
\begin{equation}
(\mathsf{M})_{ij} = \delta_{ij} + \Delta t \hat{K}_{ij} 
\; ,
\end{equation}
with $\hat{K}_{ij}$ being the Fourier representation of $K_{ij}$, the
corresponding matrix is denoted as $\hat{\mathsf{K}}$ below.
This semi-implicit method yields a much higher level 
of numerical stability than an Euler
integration. However, both integration methods are formally accurate to
$\mathcal{O}(\Delta t)$, meaning that larger $\Delta t$ values should still be
used with caution. For our particular model of consideration, we can introduce
the field basis $\phi_i = (w, \psi)_i$. A quadratic expansion of the field
Hamiltonian in Eq.~\ref{eq:H_field} can then be shown to give
\begin{subequations} \label{eq:K_matr}
\begin{align}
\hat{\mathsf{K}}(k \neq 0) &= \begin{pmatrix}
\rho \hat{\Gamma}^2_{k} \gmm{k} + 1/v & \rho \hat{\Gamma}^2_{k} \gmc{k} \\
\rho \hat{\Gamma}^2_{k} \gmc{k} & \rho \hat{\Gamma}^2_{k} \gcc{k} + k^2/4 \pi \lB
\end{pmatrix} \, , \label{eq:K_k_not_zero} \\
\hat{\mathsf{K}}(k = 0) &= \begin{pmatrix}
1/v & 0 \\
0 & 0
\end{pmatrix}
\, ,
\end{align}
\end{subequations}
where $k$ is wave number as before, 
$\rho \equiv \np N / V$ is the bulk density of polymer beads, 
$\hat{\Gamma}_k = \exp(-\bar{a}^2 k^2 / 2)$ is the Fourier transform
of $\Gamma(\rr)$, and
\begin{subequations}
\label{eq:K2}
\begin{align}
\gmm{k} &= \frac{1}{N} \sum_{\alpha,\gamma=1}^N \e^{- |\alpha -\gamma | b^2 k^2 / 6 } \, ,  \\
\gmc{k} &= \frac{1}{N} \sum_{\alpha,\gamma=1}^N \sigma_\alpha \e^{- | \alpha - \gamma | b^2 k^2 / 6 } \, , \\
\gcc{k} &= \frac{1}{N} \sum_{\alpha,\gamma=1}^N \sigma_\alpha \sigma_\gamma \e^{- |\alpha - \gamma | b^2 k^2 / 6 }  \, 
\end{align}
\end{subequations}
follow from the expansion of the term $-\np \ln Q$ in Eq.~\ref{eq:H_field}. 
The linear part of the field Hamiltonian expansion for our model 
in Eq.~\ref{eq:Field_hamiltonian_expansion} is $a_i = ( - \ii N \np, 0)_i$.
\\

\noindent
{\bf 3.3 Computational implementation of FTS}

We are now in a position to
summarize how FTS is done in a computer. First, we discretize the periodic
cubic box of volume $V$ into $M^3$ voxels, each with volume $\Delta V = V/M^3$
(Fig.~4a). The continuous fields are now approximated by
specifying their values on each voxels.  Next, we follow the steps listed below
to self-consistently evaluate the equilibrium evolution of the system dictated
by the interaction Hamiltonian Eq.~\ref{eq:H_field}.  
\begin{enumerate}
\item Initialize the fields on each voxel by complex random numbers at the first CL step.
\item Compute the system-specific $\hat{\mathsf{K}}$ 
matrix necessary for the semi-implicit method 
(Eqs.~\ref{eq:K_matr} and \ref{eq:K2}).
\item Calculate forward and backward propagators, single-chain partition 
function, charge and bead density operators 
using Eqs.~\ref{eq:propagators} and \ref{eq:Q_rhos}.
\item Update the field values on each voxels using Eq.~\ref{eq:semi_implicit_update}.
\item Start the next CL step by repeating steps from step 3 with the new fields.
\end{enumerate}
Equilibration of the system may be monitored, e.g., by plotting $Q$ vs CL time.
Once the system has stabilized, averages of operators can be calculated using
Eq.~\ref{eq:averages}. However, field configurations used in averaging should
be separated by a sufficient number of CL steps in between to make sure the
configurations are uncorrelated. The equilibration of a model system
is illustrated in 
Figs.~4b and c,
which show snapshots of the real, non-negative part of
the bead density field operator $\tilde{\rho}(\rr)$ under single-phase and
phase-separating conditions, respectively. These snapshots were generated using
the FTS python script available at {\footnotesize
\href{https://github.com/jwessen/IDP_phase_separation/blob/main/FTS_polyampholytes.py}
{\tt
https://github.com/jwessen/IDP{\textunderscore}phase{\textunderscore}separation/blob/main/FTS{\textunderscore}polyampholytes.py}}.
\\

\begin{figure}[ht!]
 \centering
  \includegraphics[width=0.80\textwidth]{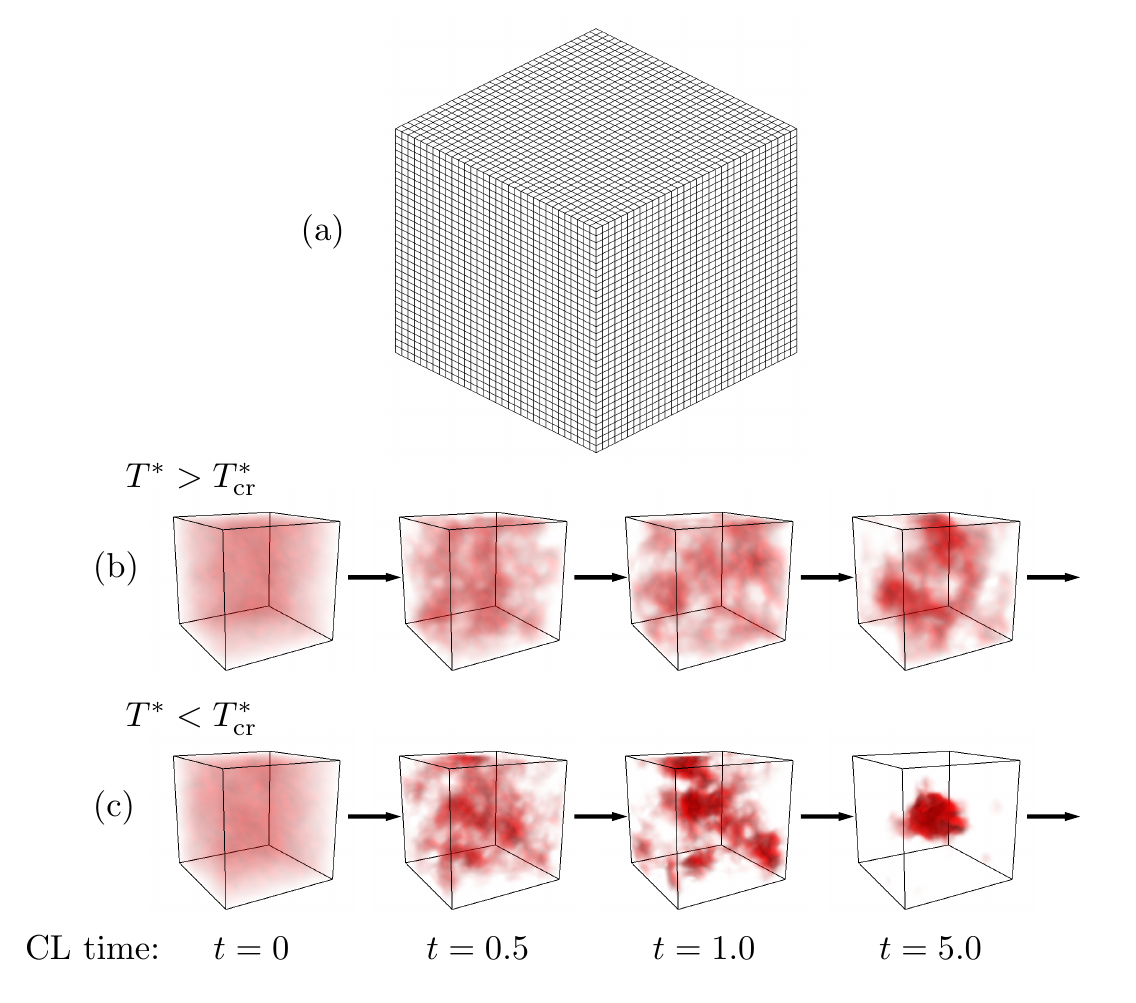}
 \caption{FTS by CL sampling.
(a) Simulation grid. (b) The initial equilibration of a system during FTS,
illustrated by snapshots of the real, non-negative part of the polymer bead
density operator $\tilde{\rho}(\rr)$. The snapshots were generated by
integrating the CL evolution equations on an $M=48$ lattice with lattice
spacing $\Delta x = \bar{a}$ and time-step $\Delta t=0.01$, starting from
random initial field configurations. The system has a bulk density $\np N /
V=0.1 b^{-3}$ of sv20 chain beads (charge sequence given in Fig.~6a),
described by a smearing length $\bar{a}=b/\sqrt{6}$ and excluded volume
parameter $v = 0.0068 b^3$. The simulation was performed at a
reduced temperature $T^* \equiv b/\lB = 10.0$ 
which is sufficiently high to prevent phase separation;
$T^*_{\rm cr}$ is critical temperature. (c)
Same as (b), but for $T^*=0.1$. At this lower temperature, the system enters
into an inhomogeneous state containing a dense, phase separated droplet in a
dilute environment. } 
\label{fig:equil}
\end{figure}

\noindent
{\bf 3.4 Chemical potential and osmotic pressure}

To investigate the propensity for the system to phase separate, we first 
describe how to compute using FTS and in units of $k_{\rm B}T$
the chemical potential $\mu $ and osmotic pressure $\Pi$, defined as
\begin{subequations}
\begin{align}
\beta \mu &= - \left( \frac{\partial \ln Z}{\partial \np} \right)_V\, , \label{eq:chem_pot_def} \\
\beta \Pi &= \left( \frac{\partial \ln Z}{\partial V} \right)_{\np} \, , \label{eq:pressure_def}
\end{align}
\end{subequations}
where $\beta= 1/k_{\rm B}T$. (This $\beta$ should not be
confused with the phase label used in Sect.~2). 
Given how $\beta \mu$ and $\beta \Pi$ vary with polymer bead
density, the possibility of phase separation can then be studied as follows:
Two phases with bead bulk densities $\rho^{(1,2)} \equiv \np^{(1,2)} N /
V^{(1,2)}$ can coexist in equilibrium if their respective chemical potentials
and osmotic pressures match, i.e.~$\beta \Pi(\rho^{(1)}) = \beta
\Pi(\rho^{(2)}) $ and $\beta \mu(\rho^{(1)}) = \beta \mu(\rho^{(2)}) $, since
the balance of osmotic 
pressure ensures that the boundary between the two phases is
stationary, while the chemical potential balance means that there is no net
flow of particles across the phase boundary. The ability for a system to phase
separate at a given temperature is therefore manifested by the existence of 
self-intersection points of the curve $(\beta \Pi(\rho) , \beta \mu(\rho) )$ 
generated by varying $\rho$ at that temperature \cite{Kardar2007}.

The partition functions $Z$ in Eq.~\ref{eq:Z_part} and Eq.~\ref{eq:Z_field} can
differ by factors of the form $(\mbox{const.})^{\np}$ and $(\mbox{const.})^{V}$
where $(\mbox{const.})$ is any factor independent of $V$ and $\np$. Such
factors have no physical consequence on phase equilibria since they only
contribute as additive constants to the osmotic pressure
and chemical potential, and
will therefore always cancel in balance equations. The factors $Z_{w,\psi}$ in
Eq.~\ref{eq:Z_field} amount to such constant contributions to the osmotic
pressure. Similarly, sources of constant contributions to $\beta \mu$ include
excluded-volume and electrostatic self-energies in $\hat{H}_{1,2}$, and the
normalization factor $( 3 / 2 \pi b^2)^{3(N-1)/2}$ in
$Q[\breve{w},\breve{\psi}]$ (which is chosen for convenience to make
$Q[0,0]=1$).

Applying the definition of chemical potential in Eq.~\ref{eq:chem_pot_def} to
the field form of the partition function in Eq.~\ref{eq:Z_field} leads to
\begin{equation}
\beta \mu = \ln \frac{\np}{V} - \left\langle \ln Q[\breve{w},\breve{\psi}]  \right\rangle_{\rm F} \, .
\end{equation}
A direct application of Eq.~\ref{eq:pressure_def} leads to an osmotic pressure
operator that suffers from UV divergences. These divergences can be regulated
by subtracting the contributions from $Z_{w} Z_{\psi}$ (which can be calculated
analytically since $Z_{w}$ and $Z_{\psi}$ are exactly solvable Gaussian
integrals). However, a much more efficient approach exists where the factors
$Z_{w} Z_{\psi}$ are used to derive a regularized osmotic pressure operator
that is manifestly insensitive to the fluctuations on scales $\lesssim \bar{a}$
\cite{Villet2014}. Before taking the volume derivative in
Eq.~\ref{eq:pressure_def}, we consider the rescaled coordinates ${\bf z} =
V^{-1/3} \rr$ and write the functional integrals in terms of the rescaled
fields $w'( {\bf z} ) = V^{1/2} w(\rr) $ and $\psi'( {\bf z} )= V^{1/6}
\psi(\rr) $. In effect, this isolates the volume dependence to the single chain
partition function $Q[V^{-1/2}\breve{w}', V^{-1/6} \breve{\psi}']$.
Eq.~\ref{eq:pressure_def} then leads to 
\begin{equation}
\label{eq:pressure_operator}
\beta\Pi = \frac{\np}{V} +\left \langle \frac{\np}{Q[ V^{-1/2} \breve{w}', V^{-1/6}\breve{\psi}']}
                \frac{\partial Q[V^{-1/2}\breve{w}',  V^{-1/6} \breve{\psi}']}{\partial V}\right\rangle_{\rm F},
\end{equation}
where the field average is understood to be taken with respect to $w'$ and
$\psi'$. After computing the volume derivative of the single chain partition
function in Eq.~\ref{eq:pressure_operator}, and returning to the original
coordinates $\rr$ and field variables $w$ and $\psi$, we obtain a regularized
osmotic pressure operator as 
\begin{equation}
\beta \Pi = \frac{\np}{V} - \left\langle \frac{\np}{V} \int \frac{\d \rr}{V} \left[ \frac{b^2}{9}\mathcal{P}_{\nabla} + \sum_{\alpha=1}^{N}\mathcal{P}_{\alpha} \left\{ \ii (\Gamma _2 -\frac{\Gamma}{2}) \star w + \ii \sigma_{\alpha}(\Gamma _2 -\frac{\Gamma}{6}) \star \psi \right\} \right] \right\rangle_{\mathrm{F}}
\; ,
\end{equation}
where 
{\hbox{$\mathcal{P}_{\nabla} = Q^{-1}\sum_{\alpha=1}^{N} q_F(\rr, \alpha) 
\nabla ^2 \exp[ \ii W (\rr, \alpha )] q_B(\rr,\alpha)$}},
{\hbox{$\mathcal{P}_{\alpha} = Q^{-1} q_B(\rr,\alpha) 
\exp[ \ii W(\rr, {\alpha})] q_F(\rr,\alpha)$}}, 
and $\Gamma _2(\rr) = (1 - r^2/3\bar{a}^2) \Gamma(\rr)$ 
with Fourier transform 
$\hat{\Gamma}_2({\bf k}) = (a^2 k^2/3) \hat{\Gamma}({\bf k})$.

Figure~5 shows $\beta \Pi(\rho)$ and $\beta \mu(\rho) $ computed
using the functions in the FTS code
{\footnotesize
\href{https://github.com/jwessen/IDP_phase_separation/blob/main/FTS_polyampholytes.py}
{\tt https://github.com/jwessen/IDP{\textunderscore}phase{\textunderscore}separation/blob/main/FTS{\textunderscore}polyampholytes.py}.}
For each bulk density $\rho$, a polymer solution object \texttt{PolySol} was
created on an $M=32$ lattice with $\Delta x =\bar{a}=b/\sqrt{6}$, $v=0.0068
b^3$, $\lB = 0.38 b$ and charge sequence sv20. 
The model polyampholyte sequence sv20 is one of the thirty
overall neutral ``sv'' sequences in ref.~\cite{rohit2013}. Here we use
several ``sv'' sequences as well as the as1, as4 model polyampholyte sequences 
from ref.~\cite{DasPCCP2018}
as examples in Fig.~5 and some of the subsequent figures.
The field variables, represented
by $w \rightarrow$\texttt{w} and $\psi \rightarrow$\texttt{psi}, were initially
set to random complex values using the \texttt{PolySol.set\_fields} function.
The fields were then evolved with time-step $\Delta t=0.01$ in CL time using
the \texttt{CL\_step\_SI} function, which implements the semi-implicit
integration scheme. At every 50th step, the polymer chemical potential $\beta
\mu$ and osmotic pressure $\beta \Pi$ were computed using
\texttt{PolySol.get\_chem\_pot()} and \texttt{PolySol.get\_pressure()}. After
an initial $10^3$ steps, the thermal averages were computed during a 
period of $4 \times
10^4$ steps. Upon plotting $\beta \mu$ against $\beta \Pi$, we see that the
resulting curve self-intersects, signalling phase separation. The condensed and
dilute bulk densities are then obtained by numerically finding the
corresponding two densities of the self-intersection point. This procedure can
then be repeated at several temperatures to obtain the binodal curve enclosing
the co-existence region in a $(\rho,T^*)$ phase diagram, where
$T^*=b/\lB$ is the reduced temperature. Figure~6
shows such phase diagrams in the $(\rho,T^*)$ plane,
computed for several model sequences. This method of computing the phase
diagram gives the critical chemical potential and pressure as a by-product,
which can be used to map the coexistence curve in $(\beta \mu,T^*)$ and
$(\beta \Pi, T^*)$ phase diagrams, shown here in Fig.~7
for the sequences in Fig.~6a. 
To elucidate the relationship between FTS and approximate analytical
formulations, the FTS results in Figs.~6 and 7 (filled circles) are contrasted
with RPA predictions (solid line in Fig.~6 and dotted lines in Fig.~7)
to be described below.
\\

\begin{figure}[ht!]
 \centering
  \includegraphics[width=0.6\textwidth]{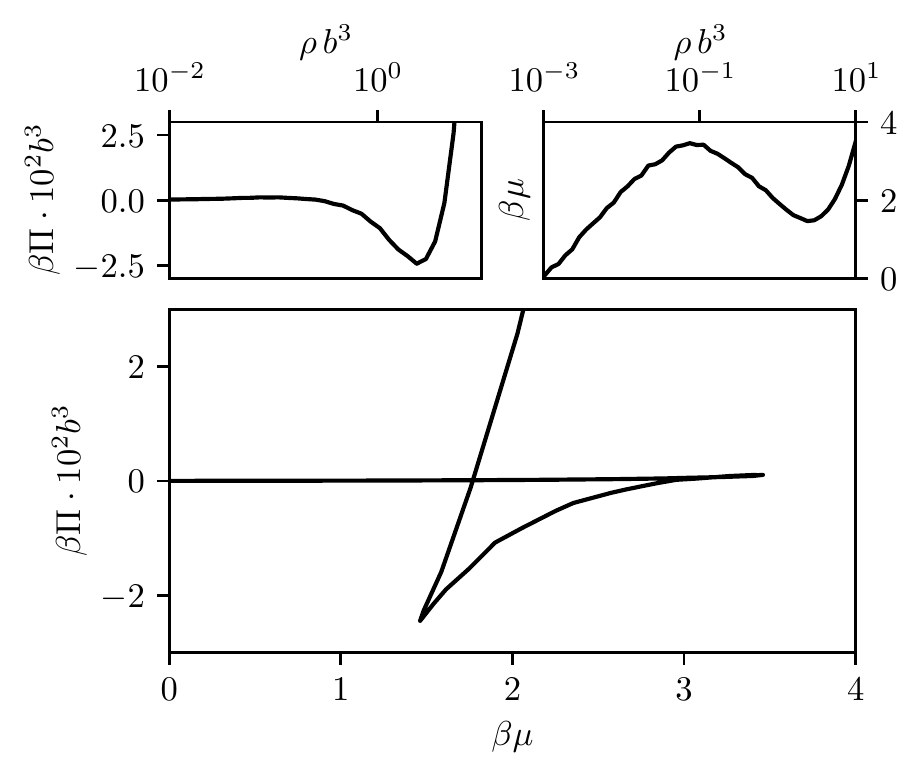}
 \caption{Chemical potential $\beta \mu$ and osmotic pressure $\beta \Pi$ 
computed in FTS in a system of sv20 chains under phase separation conditions;
$v=0.0068b^3$, $\lB=0.38b$, and $M=32$.}
 \label{fig:mu_Pi}
\end{figure}

\begin{figure}[ht!]
 \centering
  \includegraphics[width=0.7\textwidth]{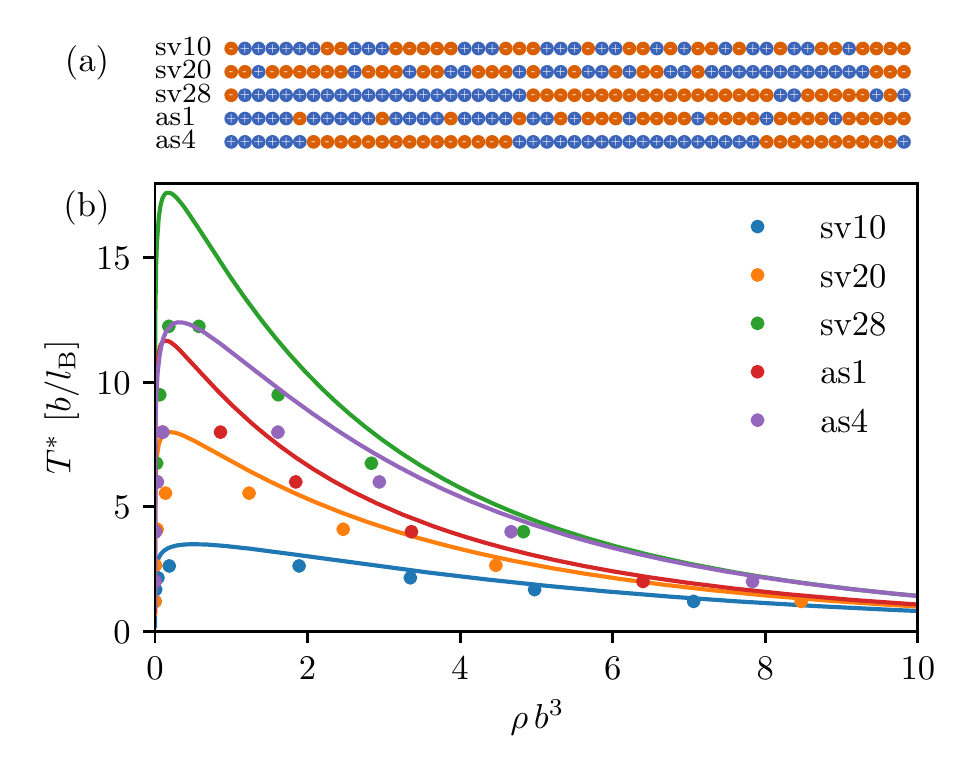}
 \caption{Comparing FTS with RPA.
(a) Model charge sequences considered in this work. Positively and negatively
charged beads are depicted, respectively, in blue and red. 
(b) $(\rho, T^*)$ phase
diagrams computed in FTS (filled circles) and RPA (solid lines);
$v=0.0068b^3$ and $M=32$ are used for FTS. The present results for
sv10 and sv20 are consistent with the corresponding results in Fig.~3C of
ref.~\cite{McCartyJPCL2019}. }
\label{fig:phase_diagrams}
$\null$
\end{figure}

\begin{figure}[ht!]
 \centering
  \includegraphics[width=0.7\textwidth]{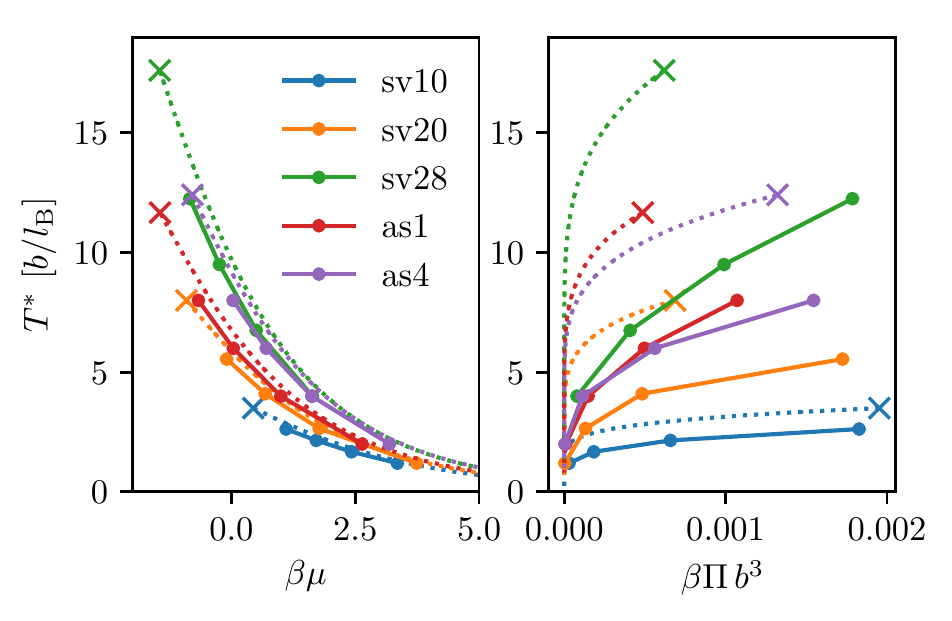}
 \caption{Chemical potential and osmotic pressure of coexisting phases.
$(\beta \mu,T^*)$ and $(\beta \Pi, T^*)$ phase diagrams, i.e., 
$\beta\mu(T^*)$ and $\beta\Pi(T^*)$ of the coexisting phases, are 
computed using FTS (filled circles) and RPA (dotted lines) for
the five model sequences in Fig.~6 (same color key). 
As in Fig.~6, $v=0.0068b^3$ and $M=32$ are used for FTS.
Crosses indicate the position of the critical point estimated in RPA.
Solid lines through the FTS data points are merely a guide for the eye.}
 \label{fig:mu_Pi_phase_diagrams}
\end{figure}

\noindent
{\bf 3.5 Random phase approximation (RPA)}

As an alternative approach, for certain applications one may opt to sacrifice
the high degree of accuracy of FTS in favor of numerical efficiency, in
which case one can compute $\beta \mu$ and $\beta \Pi$
analytically by neglecting field fluctuations beyond quadratic order; i.e.,~by
using the expanded field Hamiltonian in
Eq.~\ref{eq:Field_hamiltonian_expansion} instead of the full Hamiltonian in
Eq.~\ref{eq:H_field}. 
This approach is known as the random phase approximation (RPA). In
RPA, the functional integrals that appear in the partition function
are Gaussian and therefore
exactly solvable. For the present model, this leads to the 
following RPA free energy in units of $k_{\rm B}T$:
\begin{equation}
- \ln Z = \ln \np! - 
\np \ln V + \frac{v (N \np)^2 }{2 V} + 
\frac{V}{2} \int \frac{\d \kk}{(2 \pi)^3} \ln \left( \frac{4 \pi \lB v}{k^2} 
\det {\hat{\mathsf{K}}} \right) \, ,
\label{eq:Z_RPA}
\end{equation}
where the matrix ${\hat{\mathsf{K}}}$ is given by Eq.~\ref{eq:K_matr} 
and the factor $4 \pi \lB v / k^2$ comes from $Z_w Z_{\psi}$. 
It is instructive to note that in the $v\rightarrow 0$ limit, aside
from the overall volume factor $V$, the last
term in Eq.~\ref{eq:Z_RPA} for $-\ln Z$ corresponds to the RPA electrostatic 
free energy $f_{\rm el}$ in Eq.~3 of ref.~\cite{Lin2016} 
in the absence of the subtraction of self-energy term
${\rm Tr}(\hat{\rho}{\hat{U}}_k)$ in this equation and
the $1/k^2\rightarrow 1/k^2[1+(ka)^2]$ in ${\hat{U}}_k$
to implement a short spatial range cutoff of Coulomb interaction in
ref.~\cite{Lin2016} 
(similar $f_{\rm el}$ are used in refs.~\cite{Lin2017a,Lin2017b}).
In other words, when the trace term is eliminated 
from $f_{\rm el}$ and $1/k^2[1+(ka)^2]$ in ${\hat{U}}_k$
is reverted back to $1/k^2$ in ref.~\cite{Lin2016}, 
the last term in Eq.~\ref{eq:Z_RPA} is seen to be equal
to $Vf_{\rm el}$. This correspondence is consistent with the fact that
excluded volume is accounted for by incompressibility 
in the RPA theory of ref.~\cite{Lin2016} 
but not by an explicit term in the Hamiltonian as in Eq.~\ref{eq:Z_RPA}
and that UV divergences are regularized by the short spatial range cutoff of 
Coulomb interaction in ref.~\cite{Lin2016} rather than using the 
Gaussian smearing in Eq.~\ref{eq:Z_RPA}.

Applying Stirling's approximation $\ln \np! \approx \np \ln \np - \np$
to Eq.~\ref{eq:Z_RPA}, 
the chemical potential and osmotic pressure in RPA become
\begin{subequations}
\begin{align}
\beta \mu &= \ln \frac{\rho}{N} + v N \rho + \frac{N}{4 \pi^2} \int_0^{\infty} \d k \, k^2 \frac{A + 2 B \rho }{1 + A \rho + B \rho^2 }  
\, , \label{eq:RPA_mu} \\
\beta \Pi &= \frac{\rho}{N} + \frac{v \rho^2 }{2} + \frac{1}{4 \pi^2} \int_0^{\infty} \d k \, k^2 \left[ \frac{A \rho + 2 B \rho^2 }{1 + A \rho + B \rho^2} - \ln\left( 1 + A \rho + B \rho^2 \right) \right] \, ,  \label{eq:RPA_Pi}
\end{align}
\end{subequations}
where $\rho \equiv \np N / V$ is the polymer bead bulk density, $A \equiv \hat{\Gamma}_{k}^2 ( v \gmm{k} + 4 \pi \lB \gcc{k} / k^2)$ and $B \equiv \hat{\Gamma}_{k}^4 4 \pi \lB v ( \gmm{k} \gcc{k} - \gmc{k}^2) / k^2$. Once the integrals 
over $k$ have been performed numerically, phase separation is indicated by 
self-intersection points of the curve $(\beta \Pi(\rho) , \beta \mu(\rho) )$. 
Solid lines in Fig.~6 and dotted lines in Fig.~7 correspond to 
RPA phase diagrams computed in this way, using the RPA code 
available as
{\footnotesize
\href{https://github.com/jwessen/IDP_phase_separation/blob/main/RPA_polyampholytes.py}
{\tt https://github.com/jwessen/IDP{\textunderscore}phase{\textunderscore}separation/blob/main/RPA{\textunderscore}polyampholytes.py}}.
These figures show that RPA tends to overestimate the phase separation
propensity, largely due to an overestimation of the dilute phase chemical
potential.
\\

\noindent{\bf 3.6 Practical guide to codes for RPA and FTS with Figs.~5--7 as
examples}

We now conclude the preceding few sections on FTS by providing a 
guide to how the functions contained in the codes 
{\href{https://github.com/jwessen/IDP_phase_separation/blob/main/RPA_polyampholytes.py}
{\texttt{RPA{\_}polyampholytes.py}}}, 
{\href{https://github.com/jwessen/IDP_phase_separation/blob/main/FTS_polyampholytes.py}
{\texttt{FTS{\_}polyampholytes.py}}},
{\href{https://github.com/jwessen/IDP_phase_separation/blob/main/FTS_trajectories_MPI.py}
{\texttt{FTS{\_}trajectories{\_}MPI.py}}}, and
{\href{https://github.com/jwessen/IDP_phase_separation/blob/main/FTS_analyze_trajectories.py}
{\texttt{FTS{\_}analyze{\_}trajectories.py}}}
available from 
{\href{https://github.com/jwessen/IDP_phase_separation}
{\texttt{https://github.com/jwessen/IDP\_phase\_separation}}}
can be used to calculate phase diagrams in RPA and FTS for the polyampholyte
solutions. More details can be found in the document\\
{\href{https://github.com/jwessen/IDP_phase_separation/blob/main/rpa_fts_guide.pdf}
{\texttt{https://github.com/jwessen/IDP{\_}phase{\_}separation/blob/main/rpa{\_}fts{\_}guide.pdf}}}.

\noindent\textbf{3.6.1 RPA:} 
When executing
{\texttt{RPA{\_}polyampholytes.py}},
the script produces (i)~a plot depicting the self-intersection behavior of the
$(\beta \mu(\rho),\beta \Pi(\rho))$ curve at an example value of $\lB$, and 
(ii)~the full phase diagram (i.e.~the binodal curve) in the $(\rho,T^*)$-plane, 
the latter corresponding to the RPA curves in Fig.~6. The RPA curves in 
Fig.~7 were
generated by computing $\beta \mu$ and $\beta \Pi$ at the densities of the
binodal curve using the \texttt{PolySol{\_}RPA.calc{\_}mu{\_}Pi} function.
Users may modify parameters in the \texttt{main} function to generate their own
desired polyampholyte RPA phase diagrams. In particular, the charge sequence
($\sigma_{\alpha} \rightarrow \mathtt{seq}$) can be selected from sequences
contained in 
{\href{https://github.com/jwessen/IDP_phase_separation/blob/main/CL_seq_list.py}
{\texttt{CL{\_}seq{\_}list.py}}}. Other model parameters are represented as
$v \rightarrow \mathtt{v}$, $\lB \rightarrow \mathtt{lB}$ and $\bar{a}
\rightarrow \mathtt{a}$.

\noindent\textbf{3.6.2 FTS:} 
The accompanying FTS code
{\href{https://github.com/jwessen/IDP_phase_separation/blob/main/FTS_polyampholytes.py}
{\texttt{FTS{\_}polyampholytes.py}}}
contains the definition of a \texttt{PolySol} object, representing a field
picture description of a polyampholyte solution, which has methods necessary
for computation of chemical potential $\beta \mu$ and osmotic pressure $\beta
\Pi$ through CL evolution. Its \texttt{main} function, which
is run upon the execution of the script, contains a usage example for
generating the CL time trajectories $\beta \mu(t)$ and $\beta \Pi(t)$.

The Python script
{\href{https://github.com/jwessen/IDP_phase_separation/blob/main/FTS_trajectories_MPI.py}
{\texttt{FTS{\_}trajectories{\_}MPI.py}}}
may be used to calculate a phase diagram using a computer cluster 
where many cores are available. The script imports
{\href{https://github.com/jwessen/IDP_phase_separation/blob/main/FTS_polyampholytes.py}
{\texttt{FTS{\_}polyampholytes.py}}}
to calculate time trajectories $\beta \mu(t)$ and $\beta \Pi(t)$ at several
densities, each trajectory is assigned to a separate CPU utilising the Python
\texttt{multiprocessing} module. The resulting trajectory files can then be
analyzed using the script
{\href{https://github.com/jwessen/IDP_phase_separation/blob/main/FTS_analyze_trajectories.py}
{\texttt{FTS{\_}analyze{\_}trajectories.py}}}
that produces one file per temperature containing the $(\rho, \beta \mu, \beta
\Pi)$ data (e.g., those shown in Fig.~5) and then calculates the
self-intersection points of the $(\beta \mu(\rho), \beta \Pi(\rho))$ curves
yielding the binodal curve. The reduced temperature ($T^*$), the densities
($\rho$'s) of the co-existing
phases, their osmotic pressures and chemical potentials are written to a file
which can subsequently be used to plot FTS phase diagrams as in Fig.~6 ($\rho$
vs~$T^*$) and Fig.~7 ($\beta \mu$ vs~$T^*$ and $\beta \Pi$ vs~$T^*$).

While our FTS method is well-suited for 
computer clusters since calculations at each $\rho$ value can be 
performed in parallel, readers are encouraged to also explore a highly 
efficient Gibbs ensemble method that does not necessitate parallel 
computation~\cite{RigglemanFredrickson2010}.
\\

\noindent{\bf 3.7 Multicomponent systems}

We have so far considered simple systems with only one
polymer component. Formally, the above FTS method can be extended in a 
straightforward manner to systems with multiple components, such as
mixtures of polymers with different charge sequences \cite{Pal2021}, or
inclusions of explicit solvents and ions \cite{WessenJPCB2021}. Restricting the
interaction types to excluded volume and electrostatics while adding more
components (labeled by $p$) to the system modifies the canonical field 
Hamiltonian, with 
$\np\ln Q[\breve w,\breve{\psi}]\rightarrow\sum_q n_q \ln Q_q[\breve w,
\breve{\psi}]$ wherein the summation now runs over all components $q$,
$Q_q$ is the partition function for a single unit of component $q$, 
and $n_q$ is the number of molecules of component $q$.

Practically,
the phase behaviour of multicomponent systems can be difficult to investigate
because of the multiple chemical potentials that need to be matched at the
phase boundary (see Sect.~2.3 above). Nevertheless, some structural
information can be obtained from FTS of a phase-separated multicomponent system
in the canonical ensemble by using density-density correlation
functions, defined as \cite{Pal2021}: 
\begin{equation}
G_{p,q}(|\rr - \rr'|)=\langle \hat{\rho}_p(\rr) \hat{\rho}_q(\rr')\rangle \, ,
\end{equation}
where $\hat{\rho}_p(\rr)$ is the microscopic density of component $p$
(analogously to Eq.~\ref{eq:micro_polymer_density}). If the system is in a
single-phase homogeneous fluid state, all correlation functions 
$G_{p,q}(r)$ will approach
the product of bulk densities $\rho_p \rho_q$ at large $r$. An inhomogeneous
state, on the contrary, will show nontrivial correlations at length scales
comparable to the system size due to the inhomogeneous partitioning of the
system components. For example, if the system contains a polymer-dense
phase-separated droplet in a dilute surrounding, the polymer self-correlation
function $G_{p,p}(r)$, with $p$ serving as label for the component type, 
takes on large values at small $r$ and decreases to
$\sim 0$ for $r$ larger than the droplet size. Cross-correlation functions
$G_{p,q}(r)$ between the polymers and other components $q$ can then
reveal to what extent these components reside inside or outside the 
droplet.

The field operators corresponding to density-density correlation functions can
be obtained by adding species-specific density source terms $-\int \d \rr \,
\hat{\rho_q}(\rr) J_q(\rr)$ to the Hamiltonian before deriving the field
theory. 
As is customary in field-theoretic analyses \cite{IZ},
functional derivatives of the partition function 
with respect to~$J_q(\rr)$ in the field 
theory, followed by setting all $J_q$ to zero, then gives the desired 
field operators. Utilising an intermediate field re-definition 
(see ref.~\cite{Pal2021} for detailed derivation) gives the following 
convenient field representations of the correlation functions, 
\begin{subequations}
\begin{align}
G_{p,q}(|\rr - \rr'|)& = \langle \tilde{\rho}_p(\rr) \tilde{\rho}_q(\rr') \rangle_{\rm F} \quad \quad \quad (p \neq q) \, , \\
G_{p,p}(|\rr - \rr'|) &= \frac{\ii }{v} \langle w(\rr) \tilde{\rho}_p(\rr') \rangle_{\rm F} - \sum_{q\neq p} \langle \tilde{\rho}_p(\rr) \tilde{\rho}_q(\rr') \rangle_{\rm F} \, .
\end{align}
\end{subequations}

As a demonstration of how the correlation functions could be used to extract
information about the structural organization of multiple species in a phase
separated droplet, here we perform FTS of two pairs of sequences, sv28--sv24 and
sv28--sv9. For the results presented in this section, we used the same basic
Python script 
for single-component FTS computations
but generalized to include multiple charge sequences. A Github
link for the script is
{\footnotesize
\href{https://github.com/mmTanmoy/IDP_phase_separation/blob/main/FTS_polyampholytes_multi_species.py}
{\tt https://github.com/mmTanmoy/IDP{\textunderscore}phase{\textunderscore}separation/blob/main/FTS{\textunderscore}polyampholytes{\textunderscore}multi{\textunderscore}species.py}}.
To ensure phase separation of all the individual sequences, we chose a small
reduced temperature
$T^* = 0.2$ or, equivalently, a large Bjerrum length $l_{\rm B} = 5b$. 
Other parameters for the FTS are $M = 32$, $\Delta
x =\bar{a}=b/\sqrt{6}$, $\Delta t = 5 \times 10^{-4}$, and $v=0.068 b^3$,
and a common bulk bead density of $0.2 b^{-3}$ for both species of the 
pair of sequences.
It is noteworthy that here we used a $v$ value which is $10$ times larger 
than the one used to calculate the phase diagrams in Figs.~6 and 7. 
As observed in ref.~\cite{Pal2021}, a relatively large $v$ value is
necessary---albeit not sufficient by itself---for demixing of different 
polyampholyte species in the condensed phases, supporting the principle 
that both a significant sequence charge-pattern mismatch and a strong 
sequence-independent, generic excluded-volume repulsion are needed for 
membraneless organelle subcompartmentalization-like demixing of 
polyampholyte species in condensed droplets.
For the results presented here, a total of $40$ independent 
simulations were conducted for each parameter set and for each run, field 
configurations were
sampled at an interval of $500 \Delta t$ for a total CL time $3 \times 10^4
\Delta t$ after discarding $2 \times 10^4$ initial CL steps for equilibration.

\begin{figure}[ht!]
\centering
  \includegraphics[width=0.71\textwidth]{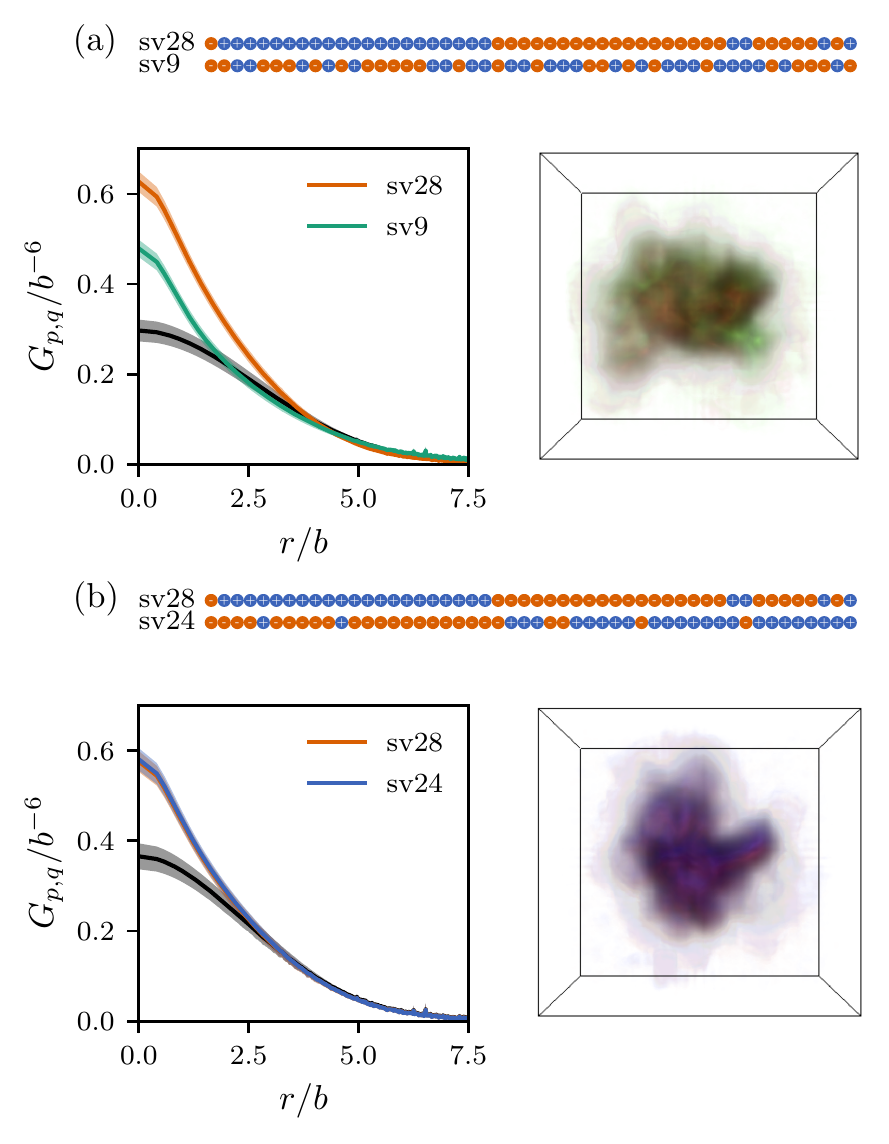}
\caption{FTS-computed density correlation and mixing/demixing of
polyampholyte species in condensed droplets.
Results are obtained using $v=0.068b^3$, $\lB=5b$, and $M=32$.
(a) Top: The charge sequences of the sv28--sv9 pair (same
color code as that in Fig.~6a). Lower-left:
Solid black line is cross-correlation function ($p\neq q$), and colored solid
lines are self-correlation functions ($p=q$) of individual sequences 
plotted using the color key shown in this panel.
The solid lines and the shaded regions around the solid lines are,
respectively, the mean and the extent of standard deviation of the correlation
functions computed over $40$ independent runs.
Lower-right: Representative snapshot of the real positive parts of the
bead-density operators $\tilde{\rho}_{\rm sv28}(\rr)$ and $\tilde{\rho}_{\rm
sv9}(\rr)$ following the color key in the left panel. 
(b) Same as (a) but for the sv28--sv24 pair.  
In this case, the self-correlation functions of sv28 and sv24 essentially
overlap. 
Data presented in this figure were generated using
a Python script which is available freely on Github through the link
{\footnotesize
\href{https://github.com/mmTanmoy/IDP_phase_separation/blob/main/FTS_polyampholytes_multi_species.py}
{\tt https://github.com/mmTanmoy/IDP{\textunderscore}phase{\textunderscore}separation/blob/main/FTS{\textunderscore}polyampholytes{\textunderscore}multi{\textunderscore}species.py}}.}
\label{fig:multi_species}
\end{figure}

The thermal-averaged correlation functions for the two pairs simulated are
shown in Fig.~\ref{fig:multi_species}. From the self-correlation function plots
and the representative snapshots, formation of a single droplet is evident.
Cross-correlation function plots demonstrate clear sequence
dependence, in the sense that the cross-correlation function for the sv28--sv9
pair peaks at a lower value compared to the sv28--sv24 pair. This feature
is interpreted as indicating that the sv28 and sv24 pair mixes more 
in the common droplet compared to sv28 and sv9 pair, as is evident
from the snapshots in Fig.~8 showing sv28 population is more concentrated in
the center of the droplet when it is paired with sv9, but is more diffused
throughout the droplet when it is paired with sv24 \cite{Pal2021}. 
The routines for computing different
correlation function could be found through the link
{\footnotesize
\href{https://github.com/mmTanmoy/IDP_phase_separation/blob/main/calc_correlations.py}
{\tt
https://github.com/mmTanmoy/IDP{\textunderscore}phase{\textunderscore}separation/blob/main/calc{\textunderscore}correlations.py}}.
In addition to the software referenced above, part of the software 
applicable to 
the FTS study of polyampholyte LLPS with explicit solvent \cite{WessenJPCB2021}
is available at
{\footnotesize
\href{https://github.com/laphysique/FTS_polyampholyte_water}
{\tt https://github.com/laphysique/FTS{\textunderscore}polyampholyte{\textunderscore}water}}.
We emphasize, however, that the code currently shared on this webpage
is only a prototype; thus it should be used with caution. 
\\

\noindent{\bf 3.8 Practical guide to multicomponent FTS with 
Fig.~8 as an example}

A step-by-step recipe for using the FTS codes for multicomponent LLPS in
the \url{https://github.com/mmTanmoy/IDP_phase_separation} repository 
is as follows. First, running the scripts requires the {\tt Python3.x}, 
{\tt Numpy}, {\tt Pandas} and {\tt Mayavi} packages. After these packages 
are installed, download the 
{\href{https://github.com/mmTanmoy/IDP_phase_separation/blob/main/FTS_Canonical.tar.gz}{\texttt{FTS\_Canonical.tar.gz}}} 
file from the 
above-cited Github repository and extract its content (untar the file).
Among its content, the 
{\href{https://github.com/mmTanmoy/IDP_phase_separation/blob/main/submit_to_cluster.py}{\texttt{submit\_to\_cluster.py}}}
script contains the \texttt{main} function that runs the simulation by 
importing the 
{\href{https://github.com/mmTanmoy/IDP_phase_separation/blob/main/FTS_polyampholytes_multi_species.py}{\texttt{FTS\_polyampholytes\_multi\_species.py}}} and
{\href{https://github.com/mmTanmoy/IDP_phase_separation/blob/main/CL_seq_list.py}
{\texttt{CL\_seq\_list.py}}} scripts. After choosing a charge sequence pair from
{\href{https://github.com/mmTanmoy/IDP_phase_separation/blob/main/CL_seq_list.py}
{\texttt{CL\_seq\_list.py}}}, 
\texttt{ncpus} copies of the \texttt{Polysol()}
object (defined in 
{\href{https://github.com/mmTanmoy/IDP_phase_separation/blob/main/FTS_polyampholytes_multi_species.py}
{\texttt{FTS\_polyampholytes\_multi\_species.py}}}) are created
and each will then be passed to the \texttt{exe} target function along with 
their unique run labels to be executed in parallel on \texttt{ncpus} CPUs. 
At this point, one can start
executing the 
{\href{https://github.com/mmTanmoy/IDP_phase_separation/blob/main/submit_to_cluster.py}
{\texttt{submit\_to\_cluster.py}}} script after setting the total CL
time steps and the CL sampling interval
(\texttt{ncpus} is a variable in
{\href{https://github.com/mmTanmoy/IDP_phase_separation/blob/main/submit_to_cluster.py}{\texttt{submit\_to\_cluster.py}}}). 
The snapshots will be saved in the
\texttt{densities} folder and the time evolution of the single chain partition
functions ($Q$s) are saved in the \texttt{evolution} folder, both on the fly.
One can observe CL time evolution of the $Q$s to determine the CL time 
needed of equilibrium. Once equilibration is achieved, subsequent snapshots
can be used as inputs to run the script 
{\href{https://github.com/mmTanmoy/IDP_phase_separation/blob/main/calc_correlations.py}{\texttt{calc\_correlations.py}}} that
output a text file containing the means and the standard deviations of the
pair distribution functions as a function of spatial distance $r$, 
which in turn can be used to compute correlation functions $G_{p,q}(r)$
such as those shown along the left column of Fig.~8. 
The script 
{\href{https://github.com/mmTanmoy/IDP_phase_separation/blob/main/plot_snapshot.py}
{\texttt{plot\_snapshot.py}}} provided under
\url{https://github.com/mmTanmoy/IDP_phase_separation}
and also in the
\texttt{visualization} folder of 
{\href{https://github.com/mmTanmoy/IDP_phase_separation/blob/main/FTS_Canonical.tar.gz}{\texttt{FTS\_Canonical.tar.gz}}}
is for visualizing the bead-density 
profiles of phase separated droplets or absence thereof by using 
the generated snapshot files as inputs.  As an example,
the snapshot files for plotting the bead-density profile along the
right column of Fig.~8 are also provided in the same folder for readers
who wish to test run the codes.
\\


\noindent{\Large\bf 4 Coarse-Grained Molecular Dynamics}\\

Because structural details are lacking in analytical 
theories and FTS, it is desirable---and often necessary---to check
their predictions against explicit-chain models that embody a higher
degree of physico-chemical realism. However, it has been a computational
challenge to apply conventional simulation techniques for 
the study of phase separation of Ising systems, relatively simple fluids, 
and homopolymer solutions \cite{Pana1992,Binder1985,Pana1987,Pana1988,Pana2000} 
to biomolecular LLPS because of the large system size the latter entails, 
the large number of chain molecules it involves, and the need to account 
for heteropolymeric sequence dependence.
Against this backdrop, a recently developed multiple-chain simulation protocol 
for coarse-grained explicit-chain IDP models emerges as an efficient 
methodology for simulating sequence-dependent IDP phase 
separation \cite{Dignon2018}.
The key advantage of this approach is that it reduces the computational 
cost significantly by initiating the main simulation with a slab-like 
configuration of condensed polymers, leading to relatively fast equilibration 
of phase-separated states by Langevin molecular dynamics \cite{Silmore2017},
although phase equilibrium can also be achieved by simulations
of similar coarse-grained IDP chain models without using the
slab-like initial state \cite{Dignon2018,koby2020}.
Since 2018, the ``slab'' methodology has been applied to study phase 
behaviors of the FUS \cite{Dignon2018,Murthy2019}, 
LAF-1 \cite{DasPNAS2020,Dignon2018}, and Ddx4 \cite{DasPNAS2020} IDRs, 
IDP-RNA mixtures \cite{Roshan2020}, IDPs
in an explicit model of water \cite{Zheng2020}, 
model polyampholytes \cite{DasPCCP2018}, mixtures of two polyampholyte
species with different sequence charge
patterns \cite{Pal2021}, as well as model polyampholytes in 
simple dipole solvent molecules \cite{WessenJPCB2021}.

In our implementation of this simulation protocol, we
have used the GPU-based HOOMD-blue package \cite{Anderson2020,Glaser2015}
version 2.5.1, which is fast computationally and user-friendly
({\href{http://glotzerlab.engin.umich.edu/hoomd-blue/}
{\tt http://glotzerlab.engin.umich.edu/hoomd-blue/}}).

As the main purpose of this chapter is to serve as a practical guide
to numerical techniques for analytical formulations,
our brief review of explicit-chain modeling of biomolecular LLPS is 
limited to this ``slab'' approach.
Other explicit-chain simulation software packages the reader
may also wish to consult---but are beyond the scope of this chapter---include 
the LAMMPS materials modeling package \cite{Plimpton1995} 
({\href{https://lammps.sandia.gov/}{\tt https://lammps.sandia.gov/}}) 
and the recently developed lattice-based 
LASSI package tailored specifically to biomolecular LLPS \cite{lassi} 
({\href{https://bio.tools/LASSI}{\tt https://bio.tools/LASSI}).
\\

\noindent{\bf 4.1 A coarse-grained explicit-chain model of IDPs}

As described originally in ref.~\cite{Dignon2018} for the ``slab'' approach, 
each amino acid residue is modeled as a single spherical bead, with differences 
in mass, electrical charge, size, and interactions to account for the 
differences among the residues. The potential energy consists of bond, 
non-bonded short-range, and electrostatic interaction terms. 
Bond-angle energies for two consecutive bonds and torsional energies 
for three consecutive bonds can be included but usually not considered. 
The bond-length terms are harmonic potentials,
with reference bond length set to the {\it trans} C$_\alpha$--C$_\alpha$
distance of 3.8 \AA.
Currently, the non-bonded interactions are mostly modeled by
either the HPS \cite{Roshan2021,Kapcha2014} or
the KH \cite{Miya1996,Kim2008} potential (energy matrix) between pairs of 
amino acid residues, which can be readily implemented using a small python 
script in HOOMD-blue. Strengths and 
limitations of the HPS and KH potentials with respect to their abilities
to capture hydrophobicity and $\pi$-related interactions have 
recently been assessed by appling these potentials to variants of 
the Ddx4 and LAF-1 IDRs \cite{DasPNAS2020}.
In addition to the interactions in the HPS and KH potentials,
other interaction types such 
as cation-$\pi$ can be incorporated~\cite{DasPNAS2020}.
Hydrophobicity scales, other than HPS, that are more
suitable for IDP simulations have also been proposed recently~\cite{Regy,RB}.
Electrostatics is tackled by the PPPM
module \cite{LeBard2012} in the HOOMD-blue package, which
has the option to include screening length to account
approximately for the effects of salt.
If the difference in salt concentration
between the dilute and condensed phases is of interest, 
it is also straightforward to model salt ions explicitly.
Details of the coarse-grained interaction schemes and their biophysical
implications can be found in refs.~\cite{DasPNAS2020} and \cite{Dignon2018}.

The coarse-grained nature of the model allows us to use a relatively
long time step of 10 fs for Langevin dynamics simulations. 
Periodic boundary conditions are usually applied to all three spatial 
dimensions. To enhance computational efficiency, we use
a cut-off distance of 15 \AA~ for non-bonded interactions.
It should be noted that HOOMD-blue uses reduced units.
Hence, one needs to convert common physical units to reduced units
before starting a simulation. This conversion process is described 
in the section on units in the HOOMD-blue package. An exposition
that uses an actual simulation system as example is also available 
in ref.~\cite{LeBard2012}.
\\


\noindent
{\bf 4.2 An efficient protocol for simulating polymer LLPS}

In the ``slab'' protocol for simulating polymer LLPS \cite{Silmore2017},
all the polymers (IDPs in our case) are initially inserted randomly in a
sufficiently large cubic box. Typically, one to several hundred IDP
chains are used.
For this initial preparation step, the Packmol package
\cite{Martinez2009} 
({\href{http://leandro.iqm.unicamp.br/m3g/packmol/home.shtml}
{\tt http://leandro.iqm.unicamp.br/m3g/packmol/home.shtml}}) or
HOOMD-blue's own module for building initial configurations can be used. 
The system is then energy minimized to remove steric clashes among molecules
using the FIRE algorithm provided in the HOOMD-blue package. 
A few hundred ps of simulation run is usually sufficient for this energy 
minimization step. It is advisable, however, to monitor the fluctuations 
in potential energy, temperature, and other thermodynamic properties 
during the course of the simulations to ensure that the run is proceeding 
properly. Once stability of the system is ascertained,
$NPT$ simulations are performed for approximately 50 ns
to compress the simulation box at a sufficient lower temperature, e.g.,
at 100 K. A low temperature is chosen for this compression process
to produce a high-IDP-density configurational state for the next step
of the simulation protocol. 
The Martyna-Tobias-Klein (MTK) thermostat and barostat \cite{Glenn1994,Mark2006}
provided in the HOOMD-blue package may be used for this $NPT$ simulation.
Alternatively, one may also use the 
{\tt update.box{\textunderscore}resize}
module to compress the box. However, as it is not known a priori how much
compression would be sufficient to produce a desirable high-density 
state, $NPT$ followed by box-resizing
is a better option per our experience. 
In this connection, it should be noted that this $NPT$ simulation step
is merely a computational technique to achieve a high-IDP-density state.
When explicit solvent is absent, the pressure applied to the simulation 
box is not equivalent to the physical hydrostatic pressure experienced by
the corresponding real-world IDP solution system. 

The compressed box is then expanded symmetrically along the direction of 
one of the box edges (referred to as the $z$-axis) 
by using the {\tt update.box{\textunderscore}resize}
module at a sufficiently low temperature.
During the expansion process, molecules are unwrapped only in the $z$-direction
because the periodic boundary conditions in the $x$ and $y$ directions
remain unchanged. The expansion process results in a slab of condensed 
IDPs in the middle of the elongated simulation box. The system is then
equilibrated by Langevin dynamics at the temperature of interest for 
2 $\mu$s. The production run is finally carried out for another
$\sim 4 \mu$s. In our experience \cite{DasPNAS2020} and that of
others \cite{Dignon2018}, an overall 6-$\mu$s simulation run is usually
sufficient for simulating the phase behavior at one given temperature
when the system contains $\sim 100$ IDP chains and each chain is
consisting of a few hundred amino acid residues.
Recognizing that the friction coefficient in Langevin dynamics affects
kinetic but not equilibrated thermodyanmic properties, weak friction 
coefficients are used in these simulations to accelerate configurational 
sampling for the computation of phase diagrams.

For systems in which model IDPs are simulated with
explicit solvents \cite{WessenJPCB2021}, an equilibrated
system comprising only of the IDPs is first prepared 
in accordance with the above procedure.
The solvent molecules are then inserted into the simulation box 
using either one's own script or the Packmol 
package \cite{Martinez2009}. During the insertion process, suitable 
distance criteria should be maintained to avoid steric overlaps. After 
inserting the solvent molecules, the system is equilibrated again 
before commencement of the production run.
\\

\noindent
{\bf 4.3 Construction of phase diagrams from simulation data}

The coexistence curve (phase diagram) of a system with a single IDP species
is constructed by first computing at different temperatures the 
distributions of IDP population in the simulation box under the constraint 
that the center of mass is fixed at $z=0$. At a given temperature,
the distribution is determined by dividing the simulation box
into many small bins along the $z$-axis and tallying IDP density
in each bin. An example of this procedure is provided in Fig.~9.
At sufficiently high temperatures (weak interaction strengths), the density 
is essentially uniform throughout the box as the system is 
in a homogeneous state (Fig.~9a).
At sufficiently low temperatures (strong interaction strengths), a region
of maximum density appears around $z=0$ with much lower densities in the
other parts of the simulation box, indicating phase separation
(Fig.~9c). These two different densities can then be mapped
onto two points on the coexistence curve for the given temperature (Fig.~9b).
After repeating this procedure for several temperatures,
the critical temperature may be estimated. 
Following ref.~\cite{Silmore2017},
the critical temperature, $T_{\rm cr}$, and the critical 
density, $\rho_{\rm cr}$, are estimated using the scaling laws
\begin{equation}
\frac{\rho_{\rm cond}(T)+\rho_{\rm dil}(T)}{2} = \rho_{\rm cr} + 
A(T_{\rm cr} - T)
\; ,
\end{equation}
\begin{equation}
\rho_{\rm cond}(T)-\rho_{\rm dil}(T) = B(1-T/T_{\rm cr})^\nu
\; ,
\end{equation}
where
$\rho_{\rm cond}(T)$ and $\rho_{\rm dil}(T)$ are, respectively, 
condensed and dilute phase density at temperature $T$,
$A$ and $B$ are fitting parameters, and $\nu$ is a critical exponent 
set to $0.325$ \cite{Silmore2017} based on the 
three-dimensional Ising model universality class \cite{Rowlinson}. 
The red coexistence curve in Fig.~9b is produced by this fitting procedure.
As further illustrations of the simulated phase behaviors of the model 
LAF-1 system in Fig.~9, snapshots of simulated configurations
generated by the VMD package~\cite{VMD1,VMD2} 
({\href{https://www.ks.uiuc.edu/Research/vmd/}
{\tt https://www.ks.uiuc.edu/Research/vmd/}}) 
are provided in Fig.~9d and f.

Additional snapshots generated using the same package are provided
in Fig.~10 to illustrate two recent applications of the ``slab'' 
simulation protocol:
Figure~10a--e shows condensed-phase demixing of two polyampholyte species 
with significantly different sequence charge patterns, supporting 
predictions by RPA \cite{Lin2017c} and FTS \cite{Pal2021} (see also Fig.~8).
Figure~10f--j shows a polyampholyte-rich droplet in
explicit solvent as observed in a recent study of dielectric effects
in polyampholyte condensates \cite{WessenJPCB2021}.
\\

\begin{figure}[ht!]
 \centering
  \includegraphics[width=0.70\textwidth]{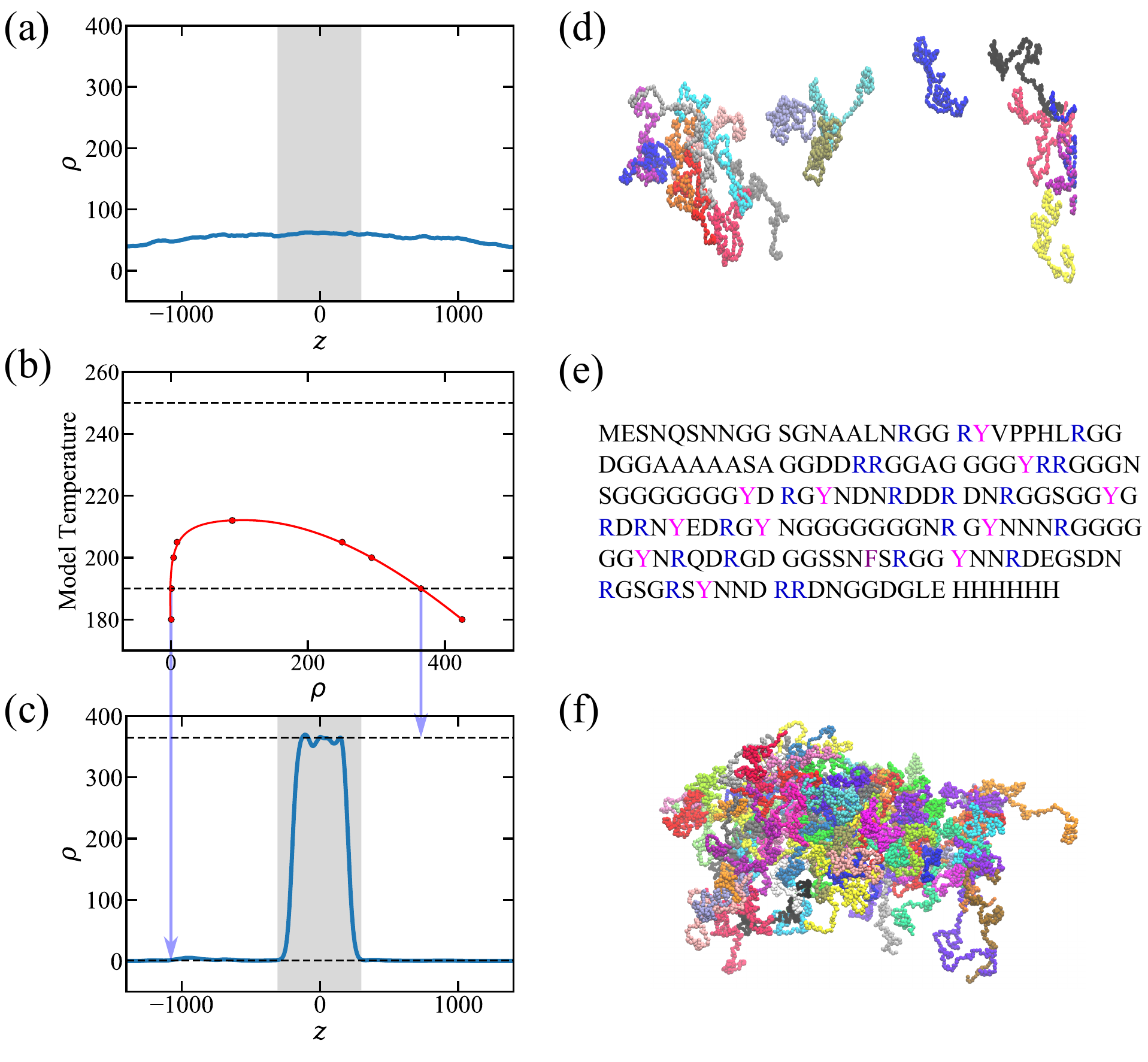}
 \caption{Coarse-grained explicit-chain simulation of heteropolymer
phase behaviors. Results shown are for the wildtype LAF-1 RGG IDR model 
studied in ref.~\cite{DasPNAS2020}. As described in this reference,
simulation is conducted by the ``slab'' protocol using 100 chains with the
KH interaction scheme \cite{Dignon2018,Kim2008}
and a uniform dielectric constant $\epsilon_{\rm r}=40$.
(a) Polymer (IDR) density profile ($\rho$, in units of mg/ml) at a
temperature indicated by the upper horizontal dashed line in (b), which is 
higher than the critical temperature of the model system. The blue curve gives
the density distribution along the $z$ dimension of the simulation box 
($z$ is in units of \AA).
(b) Phase diagram constructed from simulated data points (red circles).
All data points except the ones at model temperature = 205 are from
ref.~\cite{DasPNAS2020}. The critical point is estimated by the procedure
described in the text. The red curve is an empirical fit.
(c) Density profile at a lower temperature indicated by the lower
horizontal dashed line in (b). Phase separation is indicated by a
polymer droplet with elevated $\rho$ in the shaded region.
The light blue vertical arrows from (b) to (c) show how the points 
on the coexistence curve in (b) at a given temperature are obtained from 
the $\rho$ values in the dilute and condensed parts of the density profile 
at that temperature.
(d) A snapshot of the shaded region in (a) showing polymer chains
in a homogeneous solution. Chains are colored differently to enhance
visualization of their individual configurations.
(e) Sequence of the LAF-1 RGG IDR studied, given by the one-letter code for
the amino acids. As in ref.~\cite{DasPNAS2020}, positively charged 
arginine and aromatic tyrosine and phenylalanine residues are shown 
in blue, magenta, and deep purple,
respectively, to highlight the pattern of their positions along the sequence.  
(f) A snapshot of the shaded region in (c) showing a condensed droplet;
chains at the boundaries of the periodic simulation box are unwrapped in
this depiction.
}
 \label{fig:MD_1}
\end{figure}

\begin{figure}[ht!]
 \centering
  \includegraphics[width=0.65\textwidth]{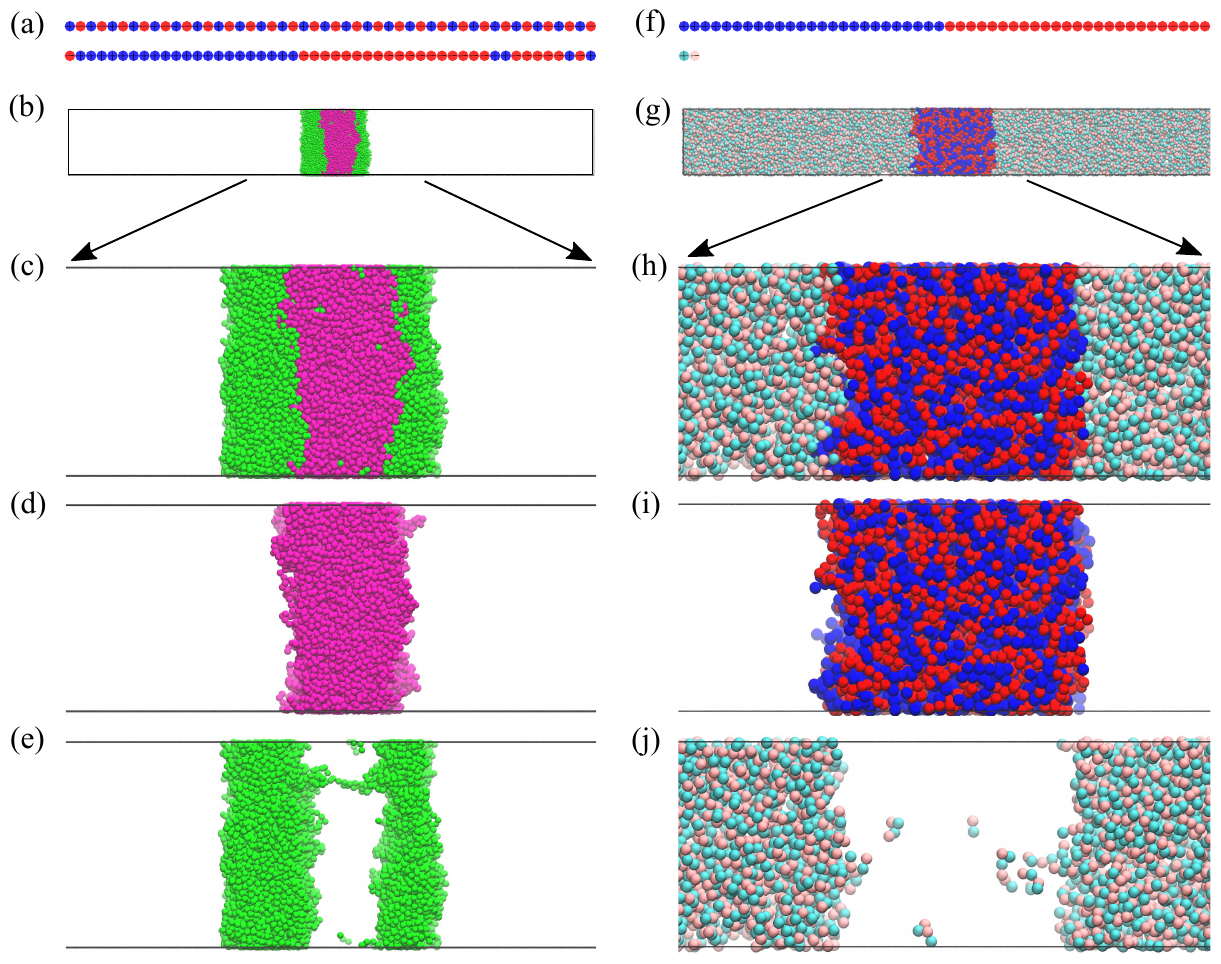}
 \caption{Condensed-phase demixing of polyampholytes
and LLPS effects of explicit polar solvent. Snapshots
from coarse-grained explicit-chain simulations using the ``slab'' protocol.
(a)--(e): A two-polyampholyte system consisting of sequences
sv1 and sv28 described in ref.~\cite{Pal2021}.
(a) sv1 (upper) and sv28 (lower), shown in the style of
Fig.~6a. (b) A snapshot of the equilibrated simulation box at reduced
temperature $T^*=0.6$; sv1 and sv28 chains are depicted, respectively,
in green and magenta.
(c)--(e): Zoomed-in view of the condensed droplet. The separate depictions
of sv28 (d) and sv1 (e) underscore that the two sequence populations 
are largely demixed in the condensed phase.
(f)-(j): A system consisting of sv30 chains and dipole solvent molecules as
described in ref.~\cite{WessenJPCB2021}.
(f) sv30 (upper, same style as (a)) and the dipole solvent molecule, for which
the positively and negatively charged beads are depicted in cyan and pink
respectively.
(g) A snapshot of part of the equilibrated simulation box at reduced
temperature $T^*=3.0$, with charges $q_{\rm d}=\pm 3.888$ on each dipole.
The color scheme of the sv30 chains and the dipoles is the same as 
that in (f) to highlight the positions of opposite charges
in the system.
(h)--(j): Zoomed-in view of the condensed droplet.
The separate depictions of sv30 (i) and dipole (j) indicate
that some dipole solvent molecules are present inside the polymer-rich
condensed droplet.
}
 \label{fig:MD_2}
\end{figure}

\noindent
{\bf 4.4 Practical guide to using the ``slab'' protocol to simulate
biomolecular LLPS}

The protocol begins with the generation of
the initial configuration of the system. This task can be
accomplished efficiently by Packmol~\cite{Martinez2009},
which offers a variety of input and output file
formats. The input script is user friendly and available from the website
(see Sect.~4.2 above). After
successful compilation of the initial configuration, the python input
script of the HOOMD-blue package is utilized for running the simulations. 
Readers who are not familiar with the input scripts for HOOMD-blue 
may first consult the examples in the tutorial section of the HOOMD-blue
website. However, while the examples on the HOOMD-blue website convey
a clear idea about how to simulate simple Lennard-Jones polymers, 
simulations of biomolecular LLPS often require more complex input scripts,
a good example of the latter is available at
{\href{https://bitbucket.org/jeetain/hoomd_slab_builder/src/master/}
{\texttt{https://bitbucket.org/jeetain/hoomd\_slab\_builder/src/master/}}},
which provides the input codes for simulating FUS 
condensates~\cite{Roshan2021}.
Schematically, the entire ``slab'' protocol may be recapitulated as follows:
(i)~Initial configuration generated from other sources such as 
Packmol + HOOMD-blue input script $\rightarrow$
(ii)~Energy minimization
$\rightarrow$
(iii)~$NPT$ box compression
$\rightarrow$
(iv)~Box expansion along $z$ axis
$\rightarrow$
(v)~Box equilibration
$\rightarrow$
(vi)~Final production run
$\rightarrow$
(vii)~Number density calculation + other analyses
$\rightarrow$
(viii)~Construction of phase diagram.
\\

$\null$

\noindent
{\Large\bf 5 Concluding Remarks}\\

In summary, we have provided a brief summary of the formulations
for several analytical theories and a practical guide to general numerical 
techniques that are useful for extracting LLPS information from these
theories in the form of two-component (one polymer solute) and three-component
(two polymer solutes) phase diagrams by determining the pertinent spinodal
regions and coexisting (binodal) conditions, using data from FH and
rG-RPA theories as examples.
With links to corresponding software, we have also presented algorithms
for performing CL sampling in FTS, which relies quite heavily on 
analytical formulation and involves considerable computation, but is
valuable for modeling sequence-dependent LLPS without invoking some of the
approximations in analytical theories such as RPA.
In addition, an efficient simulation protocol for a class of
coarse-grained explicit-chain models of biomolecular LLPS
is briefly outlined as well. The versatile methodology is important in 
its own right and also as a model with relatively more physical realism to 
compare against predictions by abstract analytical approaches such as 
FH, RPA, rG-RPA, and FTS.
While we have covered only a fraction of the theoretical/computational
approaches to biomolecular condensates, it is our hope that as a practical
guide, this chapter would be useful for researchers in the field, especially
those who have not begun but wish to initiate theoretical investigations. 
Despite considerable advances made in the past few years, 
theoretical/computational investigation of biomolecular condensates is 
still in its early stages of development.  It will take many complementary
methodologies to decipher the physical basis of the myriad biological 
functions that are being discovered by the rapidly expanding experimental 
effort on biomolecular condensates.
\\

$\null$\\
{\large\bf Acknowledgments}\\ 
This work was supported by Canadian Institutes of Health Research grant
NJT-155930 and Natural Sciences and Engineering Research 
Council of Canada grant RGPIN-2018-04351 as well as computational 
resources provided generously by Compute/Calcul Canada to H.S.C.
\\

\noindent
Y.-H.L., J.W., and T.P. contributed equally to this work.

\vfill\eject

\noindent
{\Large\bf References}\\
 \vfill\eject

\vfill\eject
\end{document}